\documentclass[letterpaper,twocolumn,10pt]{article}
 \usepackage[available]{usenixbadges}
\usepackage{usenix2019_v3}

\usepackage{amsmath}
\usepackage{mathptmx} 
\usepackage{amssymb}
\usepackage{fancyhdr}
\usepackage[normalem]{ulem}
\usepackage{url}
\usepackage{tcolorbox}
\usepackage{soul}
\usepackage{paralist}
\usepackage{algorithm}
\usepackage[noend]{algpseudocode}
\usepackage{enumitem}
\usepackage{multirow}
\usepackage{tikz}
\usepackage{soul}
\usepackage{caption}
\usepackage{stfloats}
\usepackage{enumitem}
\usepackage{amsfonts}
\usepackage{bbding}
\usepackage{algorithm}
\usepackage{url}
\usepackage{soul}
\usepackage{wrapfig}
\usepackage{tablefootnote}
\usepackage{comment}
\usepackage{caption}
\usepackage[flushleft]{threeparttable}
\usepackage{xcolor}
\usepackage[export]{adjustbox}
\usepackage{subcaption} 

\newcommand*\circled[1]{\tikz[baseline=(char.base)]{\node[shape=circle,fill,inner sep=0.5pt] (char) {\textcolor{white}{#1}};}}

\newcommand{\mirage}{\textsf{MIRAGE}}
\newcommand{\scatter}{\textsf{ScatterCache}}
\newcommand{\ceaser}{\textsf{CEASER}}
\newcommand{\ceasers}{\textsf{CEASER-S}}
\newcommand{\sass}{\textsf{SassCache}}
\newcommand{\spec}{\textsf{SPEC2017}}

\newcommand{\randomRP}{\textsf{RandomRP}}
\newcommand{\treePLRURP}{\textsf{TreePLRURP}}
\newcommand{\weightedLRURP}{\textsf{WeightedLRURP}}
\newcommand{\rriprp}{\textsf{RRIPRP}}
\newcommand{\fiforp}{\textsf{FIFORP}}

\newcommand*\filledDelta{\tikz{\draw[fill=blue, draw=blue] (0,0.03) -- (0.06,0.15) -- (0.12,0.03) -- cycle;}}

\definecolor{lightgray}{HTML}{F7F7F7}
\definecolor{customblack}{HTML}{A0A0A0}

\begin{document}

\date{}

\title{\Large \bf Systematic Evaluation of Randomized Cache Designs against Cache Occupancy}

\author{
{\rm Anirban Chakraborty$^*$}\\
Max Planck Institute for Security and Privacy\\
\href{mailto:anirban.chakraborty@mpi-sp.org}{anirban.chakraborty@mpi-sp.org}
\and
{\rm Nimish Mishra$^*$}\\
Indian Institute of Technology Kharagpur    \\
\href{mailto:nimish.mishra@kgpian.iitkgp.ac.in}{nimish.mishra@kgpian.iitkgp.ac.in}
\and
{\rm Sayandeep Saha}\\
Indian Institute of Technology Bombay    \\
\href{mailto:sayandeepsaha@cse.iitb.ac.in}{ sayandeepsaha@cse.iitb.ac.in}
\and
{\rm Sarani Bhattacharya}\\
Indian Institute of Technology Kharagpur    \\
\href{mailto:sarani@cse.iitkgp.ac.in}
{sarani@cse.iitkgp.ac.in}
\and
{\rm Debdeep Mukhopadhyay}\\
Indian Institute of Technology Kharagpur    \\
\href{mailto:debdeep.mukhopadhyay@gmail.com}{debdeep@cse.iitkgp.ac.in}
}

\maketitle
\def\thefootnote{*}\footnotetext{Equal Contribution.}\def\thefootnote{\arabic{footnote}}

\begin{abstract}

Randomizing the address-to-set mapping and partitioning of the cache has been shown to be an effective mechanism in designing secured caches. Several designs have been proposed on a variety of rationales: \textcircled{1} randomized design, \textcircled{2} randomized-and-partitioned design, and \textcircled{3} psuedo-fully associative design.
This work fills in a crucial gap in current literature on randomized caches: currently most randomized cache designs defend only contention-based attacks, and leave out considerations of cache occupancy. We perform a systematic evaluation of $5$ randomized cache designs- \ceaser, \ceasers, \mirage, \scatter, and \sass\ against cache occupancy wrt. \textit{both} performance as well as security. 

With respect to performance, we first establish that benchmarking strategies used by contemporary designs are unsuitable for a \textit{fair} evaluation (because of differing cache configurations, choice of benchmarking suites, additional implementation-specific assumptions). We thus propose a uniform benchmarking strategy, which allows us to perform a fair and comparative analysis across all designs under various replacement policies. Likewise, with respect to security against cache occupancy attacks, we evaluate the cache designs against various threat assumptions: \textcircled{1} covert channels, \textcircled{2} process fingerprinting, and \textcircled{3} AES key recovery (to the best of our knowledge, this work is the first to demonstrate full AES key recovery on a randomized cache design using cache occupancy attack). Our results establish the need to \textit{also} consider cache occupancy side-channel in randomized cache design considerations.

\end{abstract}

\color{black}
\section{Introduction}~\label{sec:intro}
Caches hide the overall memory access latency by keeping
data and instructions close to the processor. It leverages
the principle of locality of reference by buffering recently
used data such that for future references those data can be
directly served to the processor, saving multiple clock cycles.
Modern processors employ cache memories in slices, sets and ways and are typically based on set-associative design. Depending on the cache hierarchy, either the virtual or the physical address is used to map the incoming memory block into one of the ways of a particular set in the cache. In commercial processors from Intel and AMD, the address to set mapping in the last level cache (LLC) is typically done using \emph{complex addressing}~\cite{maurice2015reverse} on the virtual address. 
Due to the limited size of the cache, multiple memory blocks can be allocated to the same set, thereby creating \emph{conflicts}. These conflicting addresses that are mapped to the same cache set are called \emph{congruent} to each other. A number of works in literature exploit this phenomenon to devise a class of cache attacks called \emph{contention-based} attacks~\cite{osvik2006cache,hund2013practical,maurice2017hello,ristenpart2009hey,irazoqui2015s,yarom2014flush,zhang2011homealone,disselkoen2017prime+,liu2015last,oren2015spy,gulmezouglu2015faster,genkin2017may}. In such attacks, the attacker tries to fill up particular sets in the cache and then monitors those sets to ascertain whether targeted victim address is allocated to the same set. While other forms of cache attacks exist in literature, the attention garnered by contention-based attacks is overwhelming.

The \texttt{Prime+Probe}~\cite{osvik2006cache, liu2015last, aciiccmez2007yet, aciiccmez2010new, irazoqui2015s} is the most notable among contention based cache attacks. Here the attacker tries to create a set of congruent addresses, called \emph{eviction set}, that will potentially contend with the victim's address for a particular cache set. The success of the attack depends greatly on the efficiency of eviction set generation by the attacker for an unknown target address. In the last few years, multiple works~\cite{qureshi2019new, vila2019theory, liu2015last, bourgeat2020casa} have proposed efficient creation of eviction sets to increase the success rate of contention-based attacks. Consequently, protecting the cache, especially the LLC, as it is shared among all the cores, against such contention-based attacks has been one of the widely researched directions. A majority of cache protection schemes attempt to scramble the address-to-cache-set mapping by using randomization techniques.  Although different designs employ randomization in different ways, the broad idea is to use cryptographic primitives like block ciphers, hash functions, etc., to randomize the mappings in hardware with secret keys.

\textcolor{black}{Contemporary randomized cache schemes can be classified into three categories~\cite{purnal2021systematic,chakraborty2023randomized,saileshwar2021mirage} - \begin{inparaenum}[(i)]
\item \emph{randomized} (e.g. \ceaser),
\item \emph{randomized and partitioned} (e.g. \ceasers, \scatter, \sass) and
\item \emph{pseudo fully associative} (e.g. \mirage). 
\end{inparaenum}
} 
With the sole exception of \sass~\cite{giner2022scatter}, the other designs of randomized caches have been subjected to rigorous security evaluations~\cite{purnal2021systematic,song2021randomized,saileshwar2021mirage,chakraborty2023randomized} but only from the perspective of a particular attack, the contention-based attack. In other words, the security guarantees of these ingenious cache designs have been envisaged on the basis of the relative degree of resilience they present against the formation of eviction set. To the best of our knowledge, there has been no work in literature that evaluates these schemes with respect to other important cache attack variants\footnote{Only \sass~defends against occupancy attacks by cryptographically limiting the total number of cache lines available for attacker occupancy.}. In this work, we take an orthogonal approach to explore the security and performance properties of various state-of-the-art randomized cache designs with respect to an important but hugely overlooked cache attack - \emph{cache occupancy attacks}~\cite{shusterman2019robust, maurice2015c5, ramkrishnan2024non}.

\subsection{Motivation}
Although the number of proposed secured cache schemes and papers that claim to break such schemes try to outnumber each other, an important observation at this juncture is that most of these schemes focus on thwarting only contention-based attacks. The main rationale behind such schemes is that if address-to-set mapping can be made unpredictable and unobservable, the probability of creating an effective eviction set becomes negligible. In this pursuit, modern state-of-the-art cache protection schemes have incorporated complex structures (like specialized block ciphers, decoupling tag and data stores, localized randomization, etc.) both in software and hardware that beget the following questions: 

\begin{itemize} \itemsep0em
\item How do such schemes, engineered to be resilient against contention-based attacks, impact the performance across various replacement policies? This question becomes especially important because for an uniformly sampled key, the randomizing function has a uniform distribution over \textit{all} sets in the cache, for each cache-line install. Because of this uniform probability distribution, a particular set has higher chances of being chosen for replacement in a randomized LLC than in baseline set-associative cache\footnote{Because in addition to dependence on cache-line address, the eviction probability in case of randomized caches is also conditioned on the randomizer key and security domain identifier.}. Furthermore, various replacement policies are proposed for baseline set-associative caches, but contemporary literature mostly compares baseline set-associative cache with random replacement or plain Least-Recently-Used.

\item In an attempt to provide protection against certain types of attacks, do state-of-the-art cache protection schemes make themselves more vulnerable to other types of attacks? Specifically, wrt. randomized caches, can cache occupancy lead to adversarial success against the following threat assumptions: \textcircled{1} covert channels, \textcircled{2} process fingerprinting, and \textcircled{3} AES key recovery. 
\end{itemize}

In this paper, we answer these questions in the context of three distinct classes of secured caches, with respect to their relative resilience and performance against cache occupancy attacks (a type of cache attack that does not rely on creating eviction sets). In the microarchitectural security research community, it is usually considered that cache occupancy attacks do not pose a significant security threat to real-world critical applications. Moreover, many randomized cache designs (e.g. \mirage~\cite{saileshwar_2021}) do not consider cache occupancy attacks in their perceived threat model. However, the realm of cache occupancy channels and its potential impact on the security of randomized caches has not been thoroughly investigated in literature. In this work, we show that cache occupancy attacks can be used to perform key recovery attacks on AES T-tables. 

\subsection{Contribution}
We make the following contributions in this paper:
\begin{itemize} [leftmargin=*]\itemsep0em
\item We provide a detailed performance analysis of different randomized cache designs across multiple replacement policies and spurious cache occupancy.
\item We provide a comprehensive evaluation of multiple secured cache schemes from different design families, namely \ceaser, \ceasers, \scatter\ and \mirage\ with respect to cache occupancy attack.  This work delves into the implications of certain design principles and implementational decisions from an orthogonal perspective that has been largely overlooked in literature. 
\item We compare the cache candidates based on their relative ease of creating covert channels and adversarial fingerprinting of different workloads of \spec\ benchmark suite. 
\item Finally, we evaluate the candidate cache designs against key recovery attacks on T-table based AES-128, and provide a comparative analysis based on the guessing entropy of key recovery in each case.

\end{itemize}

\section{Background}
\label{sec:background}

\subsection{Traditional Cache Designs}
\label{sec:caches}

The classical cache found in most modern commercial processors closely follows set-associative cache design: entire cache is divided into multiple \emph{sets}, each containing a fixed number of blocks, called \emph{ways}. The address-to-set mapping is done through a deterministic function that directly depends on a certain part of the address. This deterministic mapping makes such caches vulnerable to contention-based attacks. In modern processors, the Last-Level Cache (LLC) are typically set-associative caches split into multiple \emph{slices}. This is analogous to \emph{partitioned caches}, where the cache is logically broken into multiple \emph{partitions} and an incoming address can be mapped to one of the partitions.

\subsection{Secured Cache Designs }~\label{sec:secured_cache_designs}
The majority of the modern cache attacks focus on either performing contention by setting the cache to a known state (\texttt{Prime+Probe}~\cite{osvik2006cache}) or exploiting the hit-miss ratio for shared addresses (\texttt{Flush+Reload}~\cite{gruss2016flush}). Naturally, the countermeasures proposed limit the side-channel leakage through contention. We now summarize such design principles:

\begin{figure}[!t]
    \centering
    \includegraphics[page=4,scale=0.5, trim=10 70 10 50,clip]{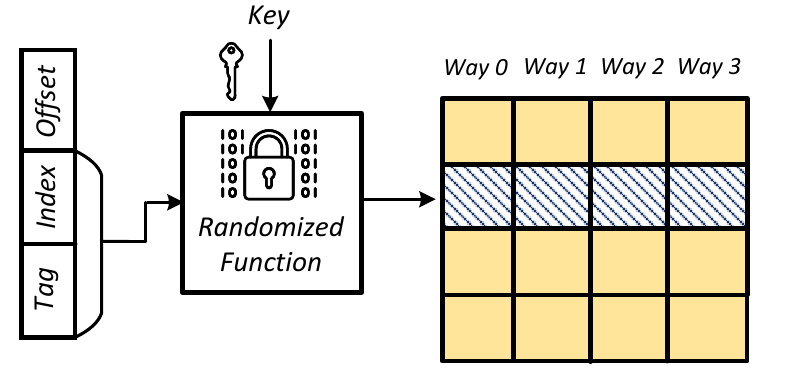}
    \caption{\centering{\ceaser\ - randomized using block cipher}}
    \label{fig:ceaser}
\end{figure}

\subsubsection{Randomized Caches}

In randomized caches, for every cache-line install, the allocated set is determined by \textit{keyed randomization} of the cache-line address by either a lookup table or a pseudo-random function (like block ciphers). Assuming the randomization key remains private, such design makes contention statistically difficult as the address-to-set mappings become unpredictable to the adversary. Earlier works like \textsf{RPCache}~\cite{wang2007new}, \textsf{NewCache}~\cite{wang2008novel} and \textsf{RandomFill Cache}~\cite{liu2014random} used permutation tables makes such designs unscalable for large LLCs.\footnote{ We thus do not focus on such designs in this work.}. More recent designs like \textsf{Time-Secured Cache}~\cite{trilla2018cache}, \ceaser~\cite{qureshi2018ceaser}, etc. alleviate the scalability issues by computing the mapping in hardware using cryptographic functions. 

\textbf{\ceaser:} \ceaser, as shown in Fig.~\ref{fig:ceaser}, uses a block cipher to convert the physical address into an \emph{encrypted address line} and uses it to index into the LLC. The encryption key is randomly generated and periodically refreshed (\emph{rekeying}) to change the address-to-set mapping after a fixed epoch period. Although block ciphers in address-to-set calculation add to the overall access latency, they are suitable for LLCs due to their large latency budget\footnote{LLCs are significantly slower compared to L1 caches.}. However, note that although the design of \ceaser~is referred to as ``randomized'', it fundamentally employs a pseudo-random mapping with a deterministic outcome~\cite{purnal2021systematic, chakraborty2023randomized}. Keeping other variables unchanged (like encryption key in a given epoch), a particular cache-line address always produces the same output.

\begin{figure}[!t]
    \centering
    \includegraphics[page=3,scale=0.48, trim=10 20 10 20,clip]{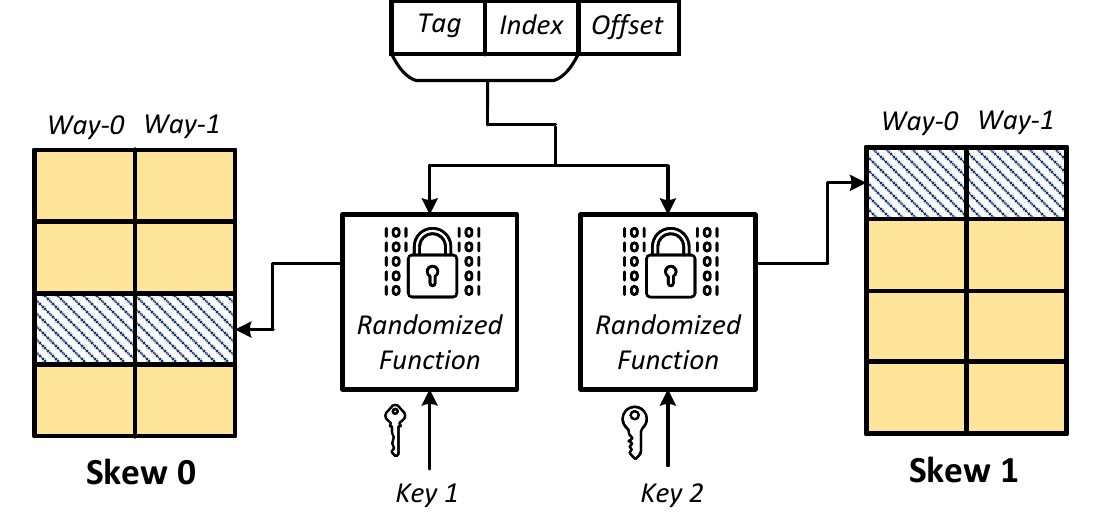}
    \caption{\ceasers - \centering{Randomized-partitioned with two keys}}
    \label{fig:ceaser-s}
\end{figure}

\subsubsection{Randomization and Partitioning} 

Another design rationale is combination of partitioning with randomization to achieve higher degree of non-determinism than randomized caches. Under this rationale, the LLC is divided into fixed number of partitions and uses a randomized mapping function to derive a different cache set in each partition, while the partition is randomly selected. The generic architecture for this type can be realized in two different ways: \textcircled{1} each partition is mapped into by using a separate key and security domain identifier, and \textcircled{2} all partitions are mapped into by using the same key and security domain identifier, but the position of the index bits is dependent on the partition number. Prominent cache designs that follow these principles are \ceasers~\cite{qureshi2019new}, \scatter~\cite{werner2019scattercache}, and \sass~\cite{giner2022scatter}.

\textbf{\ceasers:} \ceasers\ is derived from \ceaser\ by incorporating the concept of static partitioning (skewed cache). While \ceasers\ can support multiple partitions (or skews), we consider two partitions for exposition\footnote{The default \ceasers\ design contains two partitions.}. As shown in Fig.~\ref{fig:ceaser-s}, for each incoming cache-line address, every partition has its own independent instance of randomization function with a partition-specific key. The encrypted addresses thus produced are used to access both the skews concurrently. In case of a cache hit in either of the skews, the cache partition returns the data from the line that had the hit. On a miss, one of the skew is chosen randomly to install the line. The victim cache-line to replace is decided by the replacement algorithm in place.

\begin{figure}[t]
    \centering
    \includegraphics[page=1,scale=0.5, trim=10 10 10 10,clip]{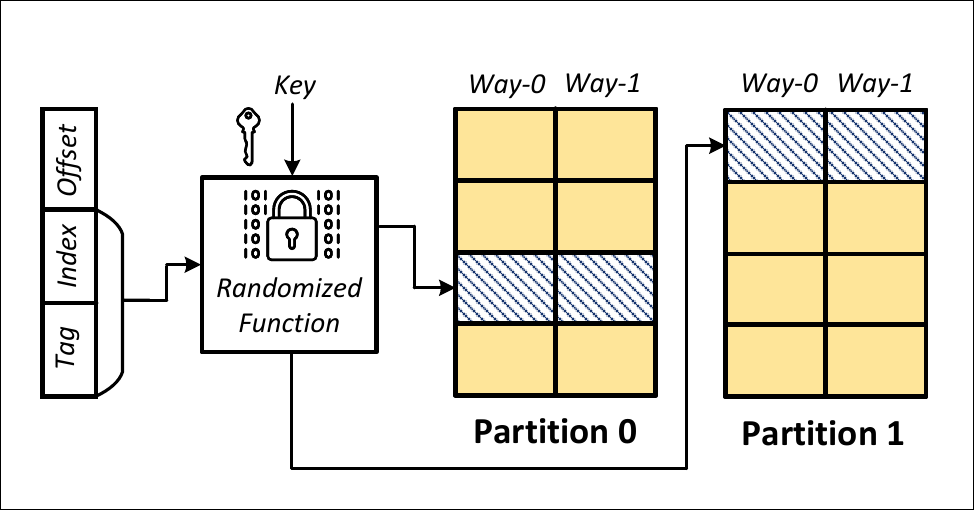}
    \caption{\textsf{ScatterCache} - randomized and skewed.}
    \label{fig:scattercache}
\end{figure}

\textbf{\scatter:} Unlike \ceasers, \scatter\ uses a single instantiation of the randomizing function with a single key. The incoming address is processed through the randomizing function, and the encrypted output is used to map into different sets in different partitions. The cache is divided into $P$ partitions (where $1 \leq P \leq$ number of ways) and the different bit-masks of the encrypted address are used to index into each partition. The choice of the partition is random. Fig.~\ref{fig:scattercache} shows a representation of \scatter\ with two partitions.

\textbf{\sass}: \sass~builds upon \scatter~and introduces keyed randomization function for \textit{each} security domain. It also ensures that the LLC occupancy for any two distinct security domains is only \textit{partially overlapping} by design. This allows \sass~to be resilient against both eviction-based attacks and occupancy-based attacks.

\subsubsection{Pseudo Fully Associative Caches}

Multiple cache designs~\cite{saileshwar2021mirage, unterluggauer2022chameleon, bhatla2024maya} have been been proposed that adopt a design rationale orthogonal to the set-associative cache indexing structure by approximating a \emph{pseudo-fully associative randomized design}. The goal of these designs is to provide security of fully associative cache while still having the practical look-up of a set-associative design.

\begin{figure}[t]
    \centering
    \includegraphics[page=2,scale=0.5, trim=10 20 10 10,clip]{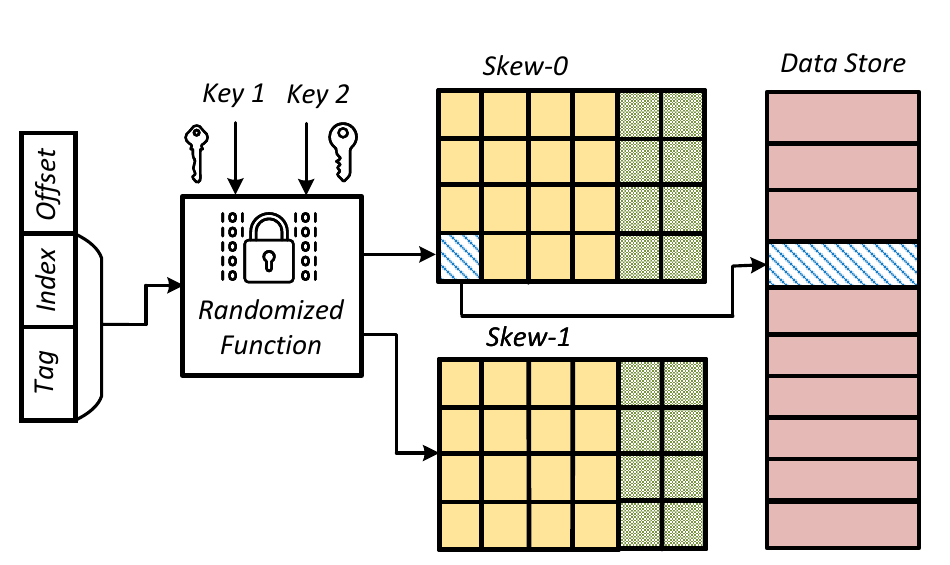}
    \caption{\mirage\ - pseudo fully associative cache.}
    \label{fig:mirage}
\end{figure}

\textbf{\mirage:} The key insight in \mirage\ is to separate the tag store and data store such that any replacement of already-installed cache-lines does not leak information about the contents of the cache. To achieve this, \mirage\ uses set-associative lookup in the tag store while maintaining full associativity in the data store. 
\mirage\ enforces \emph{global replacement} on a cache miss and provisions extra invalid tags in each set in the tag store to prevent set conflicts. Thus, when a new line is installed, an invalid tag can be allocated to it without requiring to evict an already-installed one from the same set. For the data store, a victim entry is chosen \emph{at random}, which ensures global eviction. In addition, \mirage\ also uses skewed caches with load-balancing and randomization of the input address for added security.
Fig.~\ref{fig:mirage} shows a schematic depiction of \mirage. 
 Following \mirage, a few recent schemes like \cite{unterluggauer2022chameleon} and \cite{bhatla2024maya} have been built on different variants of the pseudo fully associative design principle.

\subsection{Attacks on Secured Caches}

The classical method of generating eviction set occupies a large portion of the cache as a starting point and reduces the eviction set by eliminating addresses one at a time. This method requires $O(n^2)$ accesses~\cite{liu2015last}. Recent advances in set-eviction creation algorithms however exploit group elimination of cache-lines in $O(n)$ accesses. These attacks~\cite{qureshi2019new, vila2019theory} trivially break the early randomisation approaches like~\cite{qureshi2018ceaser}. Randomisation-with-Partition caches resist these attacks as the degree of non-determinism increases with partitioning as the address-to-set mapping depends on both the randomisation scheme and the selected skew. However, recent attacks like~\cite{purnal2021systematic, song2021randomized, chakraborty2023randomized} show faster eviction set generation techniques by observing the cache-lines evicted by the victim and probabilistically eliminating non-conflicting cachelines. 
\emph{Rekeying} is a technique that could alleviate the problem but it imposes a significant performance overhead, and the optimal key refreshment interval is difficult to determine.
In addition, some randomization designs suffer from cryptanalytic flaws due to improper implementation and design~\cite{bodduna2020brutus, purnal2021systematic}. 

\section{Performance Analysis of Randomized Cache Designs under Spurious Occupancy}
\label{sec:perf}

We now investigate and compare the performance of different randomized cache designs. While such comparisons are already performed in individual papers, there exist two main problems. First, different works consider different benchmarks, making it difficult to evaluate \textit{all} designs against a common baseline. Secondly, certain cache designs make inherent assumptions in their implementation which contemporary benchmarking strategies violate. This complicates getting a realistic view of the \textit{actual} performance of these designs.

To make a fair comparative evaluation, we begin by describing the benchmarking strategy employed by each cache design under consideration- \textcircled{1} baseline set-associative, \textcircled{2} \ceaser, \textcircled{3} \ceasers, \textcircled{4} \scatter, \textcircled{5} \mirage, and \textcircled{6} \sass. We then discuss how the implementation of \mirage~in particular makes assumptions that are not satisfied by benchmarks like \spec. We then propose a benchmarking strategy- \emph{benchmarking under spurious occupancy}- that removes this implementation assumption of \mirage~and provides a fair baseline for each cache design. Using this strategy, we provide performance evaluation of all cache designs against \spec. Finally, we also evaluate the performance of the aforementioned designs under five different replacement policies- \randomRP, \treePLRURP, \weightedLRURP, \rriprp, and \fiforp.

\begin{table}[]
\resizebox{0.5\textwidth}{!}{
\begin{tabular}{|c|c|c|c|}
\hline
Design & Platform & Benchmark  & LLC size \\
\hline
\hline
\ceaser & pin  & SPEC 2006, GAP & $8$ MB \\ 
\hline
\scatter & gem5 & GAP, MiBench, & $512$ KB \\
& &  lmbench, scimark2 & / $2$ MB \\
\hline
\scatter & \begin{tabular}[c]{@{}l@{}}Custom\\ simulator\end{tabular} & SPEC 2017 & $512$ KB / $2$ MB \\
\hline
\mirage & gem5    & SPEC 2006  & $8$ MB \\
\hline
\sass & gem5 & GAP, MiBench, & $1$ MB \\ 
& & lmbench, scimark2 & \\ \hline
\sass & \begin{tabular}[c]{@{}l@{}}Custom\\ simulator\end{tabular} & SPEC 2017 & $1$ MB \\ \hline
\hline
\end{tabular}
}
\caption{Comparison of simulation platforms and benchmark suites used to evaluate performance of different cache designs.}
\label{tab:cache_benchmarking_strategy}
\end{table}

\subsection{Contemporary Benchmarking Strategy}

Contemporary benchmarking strategy involves defining \textit{cache configurations} for a given design, and then measuring slowdown with respect to a given benchmark suite. Such cache configuration involves defining overall LLC size, number of ways, choice of randomization function (and its associated latency), L1 size, associativity, number of skews (wherever applicable), and other factors. The configurations are then simulated, and  performance statistics are recorded for the considered benchmarks. Usually, one simulation comprises a single benchmark run. In some cases (like gem5), benchmarking may involve copying the simulation over multiple cores to account for process-level fluctuations in measurements.

Table~\ref{tab:cache_benchmarking_strategy} summarizes the different cache designs, the benchmark suites evaluated, and the simulation platform. For want of conciseness, we only capture the LLC size and omit other factors considered in the respective cache configurations. It is immediately clear that a \textit{common} baseline for evaluating these cache designs is not available. We thus resort to the following benchmarking strategy (hereafter referred as  \textbf{B1}) in this work for a fair and comparative evaluation of these designs:

\noindent- \textbf{Platform}: gem5\\
- \textbf{Workload}: \spec\\
- \textbf{L1 size}: L1d : $512$ KB, L1i : $32$ KB\\
- \textbf{LLC size}: $16$ MB\\
- \textbf{LLC replacement policy} (wherever applicable): random replacement \\
- \textbf{Skews} (wherever applicable): $2$\\
- \textbf{Encryption latency}: $3$\\
- \textbf{CPU model}: TimingSimpleCPU\\

The gem5 implementation of \mirage~and \scatter~are open-sourced as part of \cite{saileshwar2021mirage}\footnote{The artifact-evaluated gem5 implementation is available at \url{http://github.com/gururaj-s/MIRAGE}.}. For~\ceaser~and \sass\footnote{The authors of \sass~generously provided us with their custom simulator for \sass, which we ported to our setting in gem5: \url{https://anonymous.4open.science/r/randomized_caches-112E/}}, we created in-house implementations in gem5. We parameterize the index derivation function such that \sass's coverage is at (recommended) $63\%$~\cite{giner2022scatter}. Finally, we consider all rate benchmarks of \spec~since they are better suited to estimating system \textit{throughput} than speed benchmarks. Throughput is directly affected by LLC performance: more LLC misses incur higher miss penalties and directly affect the throughput. Our choice of CPU model is not out-of-order, thereby giving us a clear correlation between higher LLC miss penalties and decreased system throughput.

\subsection{Assumptions on Distribution of Memory Accesses across Benchmark Lifetime}
\label{sec:mirage_assumptions}

Before we present our results, we first discuss an additional assumption in the implementation\footnote{The artifact-evaluated gem5 implementation is available at \url{http://github.com/gururaj-s/MIRAGE}.} of \mirage~which is not satisfied by general programs (including \spec). This is crucial as the said assumption has performance implications which need accounting to ensure fairness.

The goal of \mirage\cite{saileshwar2021mirage}~is to attain a negligible probability of creation of eviction sets for mounting contention based attacks. On every cache miss, \textbf{two types of evictions} are possible in \mirage\ by its specification in \cite{saileshwar2021mirage}. If the incoming address finds an invalid tag, a candidate is selected \emph{at random} from the entire data store for eviction and a \emph{reverse-pointer} based invalidation of the corresponding tag entry is performed. This is termed as \textit{global eviction} (or GLE). In contrast, if no invalid entry is found in the calculated sets in both the skews, a \emph{set associative eviction (SAE)} is performed.

However, on evaluating \mirage's artifact, we found an additional possibility for a cache-line install (other than GLE and SAE). Annotated as ``\texttt{Scenario-A}'' in the gem5 implementation, it first checks if the tag to be replaced is invalid and \emph{if there are empty slots in the datastore}. If both the conditions are satisfied, the new incoming address is put in place of the victim tag and one of the empty slots of the data cache. Moreover, to keep track of the empty slots in the large LLC data cache (which is implemented as fully-associative), the gem5 implementation uses a software-based queue named \texttt{datarepl\_vacant\_queue}. We highlight important diversions from the original design proposed in \mirage. First, this particular scenario finds no mention in the original \mirage\ paper. Secondly, use of such a software queue is not practical for LLC and is not mentioned in the original paper. In fact, authors in \mirage\ argue that designs like \textsf{RPCache}~\cite{wang2007new}, \textsf{NewCache}~\cite{liu2016newcache}, \textsf{HybCache}~\cite{dessouky2020hybcache} that perform fully-associative table-based look-ups are impractical for LLCs, whereas \mirage\ performs random replacement of data from data-cache for every \emph{invalid} tag replacement. Moving forward, we found the other two scenarios annotated as ``\texttt{Scenario-B}'' that implements SAE and ``\texttt{Scenario-C}'' that implements actual GLE were as per the original \mirage\ design, as proposed in the paper~\cite{saileshwar2021mirage}. For each benchmark in \textsf{SPEC2006}, \mirage\ fast-forwards execution by $10$ billion instructions before beginning the actual benchmarking. The intention is to allow the \textit{first} $10$ billion instructions to initialize the micro-architecture state of the LLC (i.e. exhaust all ``\texttt{Scenario-A}'' executions), post which \mirage's performance can be clearly deduced\footnote{\url{https://github.com/gururaj-s/mirage\#steps-for-gem5-evaluation}}.

However, this approach has an inherent assumption: \textit{the distribution of \textbf{load}/\textbf{store} instructions across the lifetime of every benchmark under consideration is uniform}. If this assumption is indeed true, then fast-forwarding instructions that exhaust ``Scenario-A'', and then performing the benchmarking is a sound strategy. However, if distribution of \textbf{load}/\textbf{store} across the lifetime of the benchmark is non-uniform, then \textit{using the initial portion of the benchmark to exhaust ``Scenario-A'' is an unfair strategy}. Worse, if the \textbf{load}s/\textbf{store}s are clustered during the initial stages of the benchmark, then such a strategy implies that majority of \textbf{load}s/\textbf{store}s are never evaluated under GLE (``Scenario-B'') or SAE (``Scenario-C''). Our experiments captured in Appendix~\ref{sec:mirage_assumptions_experiments} establish that this assumption is unfounded, leading us to make the following observation:

\begin{tcolorbox}[
    colback=lightgray,   
    colframe=customblack,       
    arc=2mm,             
    boxrule=1pt,          
    left=1pt,   
    right=1pt,  
    top=1pt,    
    bottom=1pt, 
]

 \textbf{Takeaway}: \mirage's gem5 implementation assumes uniform distribution of \textbf{load}s/\textbf{store}s across the lifetime of the benchmark. Not all \textsf{SPEC2017} workloads (and by extension, not all user-defined custom workloads) satisfy this criterion. \mirage~thus provides an optimistic view of the performance overhead due to global evictions (GLE).
\end{tcolorbox}

\subsection{Benchmarking under Spurious Occupancy}
\label{sec:benchmarking_spurious_occupancy}

Given the discussion in Sec.~\ref{sec:mirage_assumptions}, performing a comparative performance evaluation of all cache designs under benchmarking strategy \textbf{B1} is unfair as \textbf{B1} provides an optimistic view on \mirage's actual performance. We thus modify \textbf{B1} to a new strategy: Benchmarking under Spurious Occupancy, which adds \textit{spurious} (random) \textbf{load}s/\textbf{store}s to fill the entire LLC before the actual benchmark execution begins\footnote{For \mirage~in specific, this strategy ensures that the \textit{spurious} \textbf{load}s/\textbf{store}s are used to exhaust ``Scenario-A'', while \textit{every} \textbf{load}/\textbf{store} in the actual benchmark is vulnerable to GLE (``Scenario-C''). Strategy \textbf{B2} thus negates any advantages \mirage\ gains from ``Scenario-A'' executions.}, and ensures fairness in our evaluations. Our modified strategy (marked in \textcolor{blue}{blue}) of benchmarking under spurious occupancy (hereafter referred to as \textbf{B2}) is as follows:

\noindent- \textbf{Platform}: gem5\\
- \textbf{Workload}: \textcolor{blue}{Spurious occupancy} followed by \spec\\
- \textbf{L1 size}: L1d : $512$ KB, L1i : $32$ KB\\
- \textbf{LLC size}: $16$ MB\\
- \textbf{LLC replacement policy} (wherever applicable): random replacement \\
- \textbf{Skews} (wherever applicable): $2$\\
- \textbf{Encryption latency}: $3$\\
- \textbf{CPU model}: TimingSimpleCPU\\

\begin{figure*}[!t]
    \centering
    \includegraphics[scale=0.5]{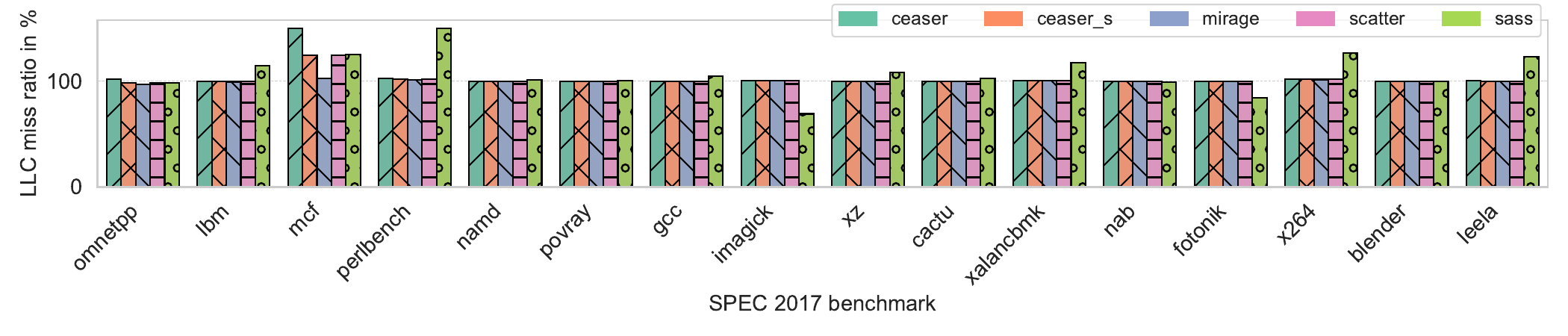}
    \caption{Performance evaluation of considered cache designs with \randomRP~replacement policy (normalized against baseline set-associative, and expressed as a $\%$). Performance statistics are averaged over $300$ copies of \spec~runs.}
    \label{fig:random_rp}
\end{figure*}

Note that \textbf{B2} is followed for \textit{all} cache design to ensure fairness. We choose the number of LLC misses as our metric for this evaluation\footnote{This is a reasonable choice since increased LLC misses directly correlate with slowdown in execution time, since each LLC miss incurs a huge miss penalty in being serviced by the memory controller.}. Fig.~\ref{fig:random_rp} captures the results of our evaluation. We follow the default replacement policy (\randomRP) for all cache designs under consideration. 

\subsection{Effect of Replacement Policies}
\label{sec:replacement_policies}

One important factor in a set-associative cache design is the choice of replacement policy invoked during set-associative eviction. Other than \randomRP, we hence conduct performance evaluation of \spec~under other replacement policies: \treePLRURP, \weightedLRURP, \rriprp, and \fiforp. Our choice of replacement policies is motivated by our intention of covering different \textit{rationales} of replacement policy design in our evaluation. For instance, \treePLRURP~and \weightedLRURP~are variants of Least Recently Used policy (LRU) that use specialized structures like binary trees and $1$-bit pointers to decide candidates for eviction. In complete contrast, \rriprp\ is a variant of \textit{not} recently used policy, and uses a re-reference prediction value to estimate whether a cache block will be used in near-future (which educates whether the replacement policy chooses the block for eviction). Finally, \fiforp~is textbook first-in, first-out replacement.

Our results are summarized in Tab.~\ref{tab:perf_eval} (we refer to Appendix Fig.~\ref{fig:treePLRURP}, Fig.~\ref{fig:weighted_treePLRURP}, Fig.~\ref{fig:rriprp}, and Fig.~\ref{fig:fiforp} for comparison at benchmark granularity)\footnote{Note that the gem5 implementation of \sass~does not support \treePLRURP~and \weightedLRURP.}. Note that in all experiments, among all designs, \mirage~has no effect of the policy because of its practically negligible occurrences of set-associative eviction. The variations observed for the the normalized score of \mirage\ stem for varying LLC misses for baseline set-associative, and not from variations in misses for \mirage. We summarize the \textbf{main takeaways}:

\smallskip

(1) For default replacement policy \randomRP, \mirage, \ceasers, and \scatter\ perform almost comparable wrt. baseline set-associative cache. 

\smallskip

(2) For \weightedLRURP and \fiforp, \ceasers~and \scatter~outperform baseline set-associative cache considerably. \weightedLRURP~and \fiforp~are the only instances where choice of a replacement policy allows randomized cache designs to \textit{outperform} baseline set-associative cache. In other words, \weightedLRURP and \fiforp~are the only instances of replacement policies where \textit{randomized} caches trump baseline set-associative caches in \textit{both} performance as well as resilience against contention based attacks.

\smallskip

(3) \mirage\ is comparable to baseline set-associative cache \textit{only} for \randomRP. Because of practically no set-associative evictions, there is no perceivable effect of changes in replacement policies on \mirage. Hence, when baseline set-associative cache takes advantage of replacement policies other than \randomRP, \mirage\ performs much worse. 

\smallskip
(4) \ceaser\ and \sass\ perform considerably worse than all their counterparts on \textit{all} evaluations.

\smallskip

(5) There is \textit{no} cache design which, for \textit{any} of the replacement policies, trumps baseline set-associative cache in \textit{all} of the following parameters: \textcircled{1} better performance, \textcircled{2} resilience against contention based cache attacks, and \textcircled{3} resilience against occupancy based attacks (cf. Sec.~\ref{sec:security}).

We attribute the poor performance of \ceaser\ to its lack of skews; this observation is helped by the fact that performance of \ceaser\ does not improve with changing replacement policies. Likewise, we attribute the poor performance of \sass\ to its design choice of security-domain based isolation of LLC occupancy which limits occupancy for a benchmark even if it has genuine requirement of LLC occupancy\footnote{As the next sections establish, it is this same design choice that allows \sass~to resist our occupancy based attack vectors, thereby establishing an interesting perspective where a design decision leads to \textit{both} worse performance but increased security wrt. contention \textit{and} occupancy attacks.}. As with \ceaser, we observe no considerable improvement in \sass\ even on varying replacement policies.

\begin{table}[]
\resizebox{0.5\textwidth}{!}{
\begin{tabular}{|c|c|c|c|c|c|}
\hline
Policy & \ceaser & \ceasers  & \mirage & \scatter & \sass \\
\hline
\randomRP & $+6.7\%$ & $+1.061\%$  & $-0.02\%$ & $+1.064\%$ & $+13.86\%$ \\ 
\hline
\treePLRURP & $+10.31\%$ & $+9.1489\%$  & $+4.00\%$ & $+9.1485\%$ & N/A \\
\hline
\weightedLRURP  & $+15.78\%$ & $-4.3028\%$  & $+5.03\%$ & $-4.3027\%$ & N/A \\
\hline
\rriprp & $+15.58\%$ & $+0.535750\%$  & $+6.42\%$ & $+0.535757\%$ & $+57.45\%$ \\
\hline
\fiforp & $+13.10\%$ & $-1.3548\%$  & $+8.22\%$ & $-1.3546\%$ & $+69.21\%$ \\
\hline
\hline
\end{tabular}
}
\caption{Comparison of different cache designs against evaluated replacement policies through reported \textit{averages of LLC misses} on \spec, normalized against baseline set-associative cache, and expressed as a percentage. Performance statistics are averaged over $300$ copies of \spec~runs. A positive $+\mathtt{x}\%$ (alt. $-\mathtt{{x}\%}$) implies the design is $\mathtt{x}\%$ slower (alt. faster) than baseline set-associative cache.}
\label{tab:perf_eval}
\end{table}

\begin{tcolorbox}[
    colback=lightgray,   
    colframe=customblack,       
    arc=2mm,             
    boxrule=1pt,          
    left=1pt,   
    right=1pt,  
    top=1pt,    
    bottom=1pt, 
]

 \textbf{Takeaway}: Benchmarking in contemporary literature is extremely diverse (wrt. choice of platform, choice of benchmarking suite, LLC configuration etc). Once we harmonize all cache designs to a uniform benchmarking strategy, we have an interesting insight: each cache design has differing affinities for replacement policies. Concretely, while most prior works are evaluated mostly with random replacement, we show that different cache designs have varied effects of replacement policies. In some cases, randomized caches actually outperform baseline set-associative cache.
\end{tcolorbox}

\section{Systematic Evaluation of Randomized Cache
Designs against Cache Occupancy Attacks}
\label{sec:security}

An important criterion for contention-based attacks like \texttt{Prime+Probe}, \texttt{Evict+Time} and similar variants, is that the attacker must fill all the ways in the target cache set to leak information about the victim.
However, a separate class of cache attacks exist in literature that do not depend on the granularity of extracted cache set information. These attacks, broadly called \emph{cache occupancy attacks}, focus on the cache footprint of the victim in terms of cache occupancy, rather than relying on activity within specific cache sets. Such attacks are powerful as they have been used in literature to perform website fingerprinting~\cite{shusterman2019robust,shusterman2020website} through javascript-supported browsers as well as covert channels for Spectre~\cite{verma2022these}. 
For victim processes that leave memory footprint in terms of cache occupancy depending on some secret bit, an attacker can recover such information simply by filling up the cache and observing the cache occupancy. Earlier attacks like \cite{cock2014last} constructed covert channels using L1 cache occupancy whereas \cite{ristenpart2009hey} showed how occupancy could even detect keystroke timing and network load in co-located virtual machines on cloud servers. \textcolor{black}{More recently, the authors in~\cite{ramkrishnan2024non} showed that cache occupancy attacks can be used to differentiate between secret keys of ciphers such as AES and RSA~\footnote{In this paper, we consider a stronger adversarial attack: key recovery on AES rather than just distinguishing keys.}.}
However, one of the major requirements, which in fact is a drawback of such an attack, is that the attacker process needs to fill up a significant portion of the cache in order to enforce  cache occupancy channel. Previous works~\cite{maurice2015c5,maurice2015reverse,lipp2020nethammer} that demonstrated covert channels and side channels using cache occupancy techniques have noisy data transfer and can be easily detected by the operating system or watchdog applications when a particular program routinely occupies a large portion of the cache.

With the sole exception of \sass, cache occupancy attacks are considered outside the threat model of state-of-the-art randomized cache designs. In this work, we reason that omitting cache occupancy attacks from design considerations of randomized caches is not prudent. Recall that design rationales for randomized caches aim to attain a uniform distribution over all LLC sets for every cache line install. While a uniform distribution of address-to-set mapping helps defending against contention attacks, on the other hand, it amplifies opportunities for a cache occupancy attack simply because every set in the LLC has a \textit{higher} probability of being selected for a cache-line install (compared to baseline set-associative cache). Formally, we ask the following question: 
\smallskip

\textit{In an attempt to achieve uniform distribution over the sets for a cache-line install, do existing randomized cache designs inadvertently end up \textbf{amplifying} cache occupancy attacks?}

\smallskip

Only \sass~claims to defend against \textit{both} contention attacks and occupancy attacks. \sass~sacrifices achieving a uniform distribution over \textit{all} sets of the LLC in favor of achieving uniformity over a \textit{subset} of the LLC. Every security domain is assigned such a \textit{subset}, and there is little overlap between two subsets. As a result, even though a given set in \sass~has a higher probability (compared to that of baseline set-associative cache) of being picked for address-to-set mapping, by not allowing the \textit{subsets} to significantly overlap, it ensures security-domain based isolation of LLC occupancy. Thereby it evades both attacks at the same time.

We evaluate the randomized cache designs under considerations under three different threat assumptions, ordered from strongest to weakest:

\begin{itemize}
    \item \textbf{Covert Channel}: This use-case has the strongest assumption: two processes \textit{collude} to covertly transmit a bit-stream over a LLC occupancy channel. It is well understood that occupancy covert channels can be created over baseline set-associative caches too. Our objective is to rather understand whether randomized caches make it \textit{easier} to do so.

    \item \textbf{Process Fingerprinting}: This use-case is a cache-occupancy based side-channel where an attacker attempts to profile victim execution. 
    
    \item \textbf{Key Recovery on AES}: This use-case has the weakest assumption and the most difficult objective: from a cache occupancy side-channel, an attacker attempts to perform key recovery of AES T-Table victim. Prior work like~\cite{ramkrishnan2024non,genkin2022cachefx} has shown cache occupancy side-channel to \textit{distinguish} between two AES keys. Full key recovery was not an objective of their attack.
\end{itemize}

Our consideration of varied threat assumptions is to establish the potency of cache occupancy attacks, as well as to demonstrate the \textit{amplification} of such attacks when combined with the design rationale of certain randomized caches.

\section{Case 1: Covert Channel}
\label{sec:covert_channel}
We consider two non-privileged processes installing addresses in the cachelines by accessing different memory locations belonging to their own address space. In line with assumptions of prior attacks on randomized caches~\cite{purnal2021systematic, song2021randomized, bourgeat2020casa, chakraborty2023randomized}, we assume the randomized LLC is shared between the two processes. We also assume that the randomization function is pseudo-random and cryptanalytically secure.

We use a Python-based functional LLC simulator for the covert channel. We configure \mirage\ with $16$MB size and $8+6$ ways in each tag set with $2$ skews. Each incoming address is randomized using \textsf{PRESENT}~\cite{bogdanov2007present} block cipher with two fixed and distinct keys. Likewise, for \scatter\ we use $16$MB cache with $2$ partitions and $4$ ways in each partition\footnote{To make the comparison fair, we configure {\sffamily\footnotesize ScatterCache} and {\sffamily\footnotesize MIRAGE} with $2$ partitions/skews and total $8$ ($4$ in each partition for {\sffamily\footnotesize ScatterCache}) ways available for each incoming address.}. Similar configurations were also maintained for \ceaser\ and \ceasers, with \ceasers\ having two partitions (skews)\footnote{We skip \sass~because its non-overlapping, security domain dependent LLC subsets make it improbable to setup the covert channel.}. We assume random replacement policy for all the cache designs.
The adversary only uses \texttt{rdtsc}-based timing measurements on its own memory accesses to determine cache hits and misses.

\subsection{Channel Setup}

We describe the channel setup to transmit a single bit. For transmitting an entire bit-string, we repeat this setup over the length of the bit-string.

\smallskip
\circled{1} \textbf{Receiver: } Allocates sufficient memory space and randomly accesses $\ell$ number of addresses. In order to ensure that each address occupies an entire cacheline, the addresses are accessed in a sequential manner with an interval of $1,000$ (without loss of generality) between each address. For example, if the first address accessed is $p$, then the next address to be accessed will be $1000 + p$. This ensures that none of the addresses experiences a cache hit.

\smallskip
\circled{2} \textbf{Sender: } The sender is assumed to be executing on a separate physical core, completely agnostic of the addresses accessed by the receiver. The sender allocates a memory range and depending on the value of the bit ($0$ or $1$) it wants to transmit, either makes $x$ or $y$ memory accesses. Additionally, the addresses accessed are at an interval of $1,000$ (wlog.) to ensure no two accesses are mapped to the same cache-line.

\smallskip
\circled{3} \textbf{Receiver: } Once the sender has completed its operation, the receiver re-accesses all the $\ell$ addresses it has previously installed in the cache. It is worth mentioning that the receiver, at this point, does not have any knowledge about the number of its own addresses being evicted from the cache. Nor does it know  the number of addresses from the sender installed presently in the cache.
The receiver re-accesses the addresses and measures the access times for each of them individually. The ones that are still in the cache will experience a cache hit, whereas the ones evicted due to contention with the sender addresses will experience a cache miss.

For synchronization between the processes, a busy wait without any memory accesses is sufficient. The sender sleeps for the time duration it takes the receiver to access $l$ addresses. The receiver sleeps for the time duration it takes the sender to access the maximum of two accesses (i.e. \texttt{max}($x$, $y$)).

\subsection{Evaluation: Different Occupancy Rates}

We perform a characterization of cache occupancy channel on all variants (under consideration) to estimate the optimal occupancy amount for each variant. We vary the number of addresses accessed by the receiver to install in the cache such that the cache occupancy rate covers the entire range of $1\%$ to $40\%$. For simplicity, we fix the number of sender accesses to $x=1000$ for transmitting `$0$' and $y=2000$ for transmitting `$1$'. The receiver only considers number of misses observed in its occupancy. For a given \% of cache occupancy, if the number of misses observed by the receiver are statistically distinguishable between transmission of `$0$' or `$1$', we consider the channel to be successfully established.

\begin{figure*}[!t]
    \centering
    \begin{subfigure}[b]{0.45\textwidth}
        \includegraphics[width=\textwidth]{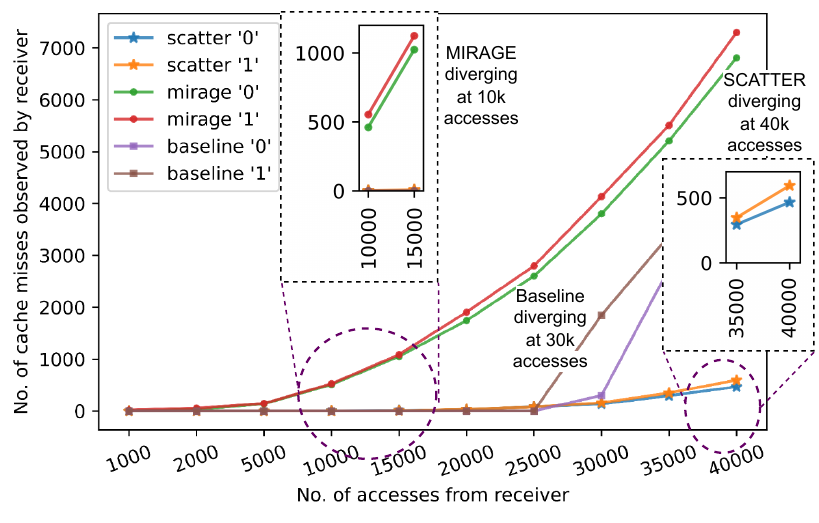}
        \caption{Comparative analysis of cache occupancy channel in \mirage, \scatter\ and baseline set-associative cache.}
        \label{fig:cache_occupancy_comparison}
    \end{subfigure}
    \hfill
    \begin{subfigure}[b]{0.45\textwidth}
        \includegraphics[width=\textwidth]{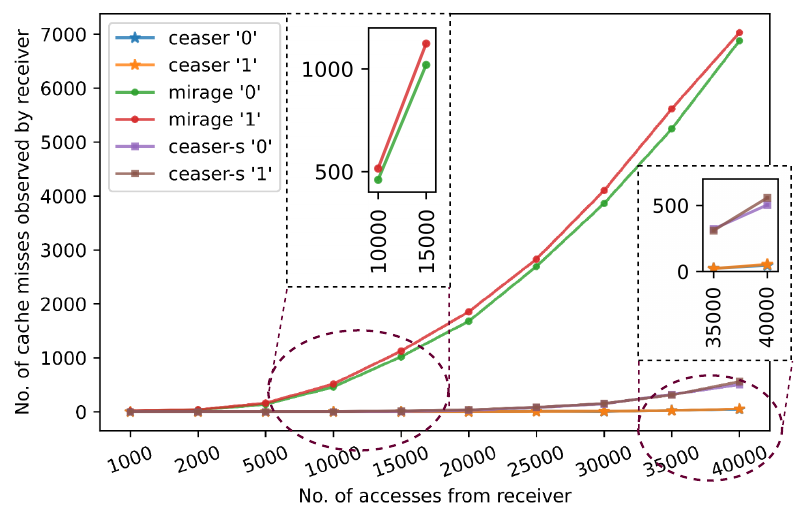}
        \caption{Comparative analysis of cache occupancy channel in \mirage, \ceaser\ and \ceasers\ cache designs.}
        \label{fig:cache_occupancy_comparison_v2}
    \end{subfigure}
    \hfill
    \caption{Comparative analysis of cache occupancy based covert channel in different cache designs.}
    \label{thr_ref_diag}
\end{figure*}

Fig.~\ref{fig:cache_occupancy_comparison} and Fig.~\ref{fig:cache_occupancy_comparison_v2} show the comparative analysis of all variants of caches under consideration with respect to cache occupancy attack on a $16$MB cache. Each data-point reported is averaged over $100$ runs of same experiment. For baseline cache, the receiver requires at least $\ell = 30,000$ accesses, while for \scatter\, it requires around $\ell = 40,000$ accesses to observe a statistically significant difference in the number of cache misses between $x$ and $y$ accesses from sender (cf. Fig.~\ref{fig:cache_occupancy_comparison}). For \ceasers, the receiver requires around $\ell = 40,000$ accesses, which is expected as the design principle of \ceasers\ is similar to \scatter; whereas, for \ceaser, the receiver requires more than $\ell = 40,000$ accesses (cf. Fig.~\ref{fig:cache_occupancy_comparison_v2}). The higher resilience of \ceaser\ could be attributed to the monolithic design approach undertaken by the designers which requires the receiver to fill up all the ways ($8$ in our case) in any set for the sender to successfully evict one of the entries. 
Furthermore,  one can observe that the number of misses corresponding to two access patterns (signifying `$0$' and `$1$') for \mirage\ start diverging at $10k$ accesses while for both baseline and \scatter\ it's much higher. Note that we were not able to observe any misses for baseline, \scatter, \ceaser\ and \ceasers\ till $25k$ accesses.


\subsection{Evaluation: Fixed \textit{low} Occupancy Rate}

It is clear than pseudo fully associative design rationale (i.e. \mirage\ ) allows a statistically distinguishable distribution of cache misses between ``$0$'' and ``$1$'' in much \textit{lesser} receiver accesses than other designs based upon set-associative design rationale. To investigate this further, we repeat the previous experiment, but with a deliberately fixed \textit{low} LLC occupancy rate. Conservatively, we fix this rate at $10\%$; refer Fig.~\ref{fig:cache_occupancy_comparison} where \mirage\ shows signs of divergence at this rate. We thus fix $10k$ accesses from the receiver and and $3k$ accesses from sender. As before, the receiver first makes $\ell = 10,000$ memory accesses, installing them in the cache and then waits for the sender to make its own accesses. All evaluation numbers are reported as averages over $100$ repetitions of the experiment.

\noindent \textbf{\textit{Observations for `0':}} To send bit `$0$', the sender makes $x = 1000$ memory accesses. For \mirage\, we observed that around $90$ addresses of the receiver that were already installed get evicted due to random replacement in the global data-store. The receiver, when re-accessing its own addresses, observes cache misses (higher latency) for around $490$ of them\footnote{Note that while the operation of the sender amounts of $90$ evictions, we observe far higher (i.e. $490$) cache misses. We attribute this to \textit{self-evictions}: for any two cache-lines $i,j \in \{1, 2, 3, \, \cdots,\, l\}: j > i$ installed by the receiver, the installation of line $j$ evicts line $i$ due to global eviction.}. Note that we did \textit{not} observe any cache misses for the receiver in \ceaser, \ceasers, \scatter\ and baseline cache.

\noindent \textbf{\textit{Observations for `1':}} To send bit `$1$', the sender makes $y = 2000$ memory accesses. In the case of \mirage\, we observed that $120$ addresses of the receiver are evicted due to this operation. The receiver now re-accesses all its $10,000$ addresses and observed cache misses for around $540$ of them. However, for other variants such as \ceaser, \ceasers, \scatter\ and baseline cache, we did not observe any misses from the receiver side.

\begin{tcolorbox}[
    colback=lightgray,   
    colframe=customblack,       
    arc=2mm,             
    boxrule=1pt,          
    left=1pt,   
    right=1pt,  
    top=1pt,    
    bottom=1pt, 
]

\textbf{Takeaway}: In the context of sender's accesses evicting a statistically significant portion of receiver's accesses, set-associative based design principles such as \ceaser, \ceasers, \scatter\ show higher resilience than fully associative cache based designs (like \mirage\ ). In other words, while \textit{all} design variants show statistically distinguishable distributions for ``$0$'' and ``$1$'', pseudo fully associative design principle requires the \textit{least} occupancy among all designs under consideration, making it more vulnerable than baseline set associative cache wrt. occupancy attacks.

\end{tcolorbox}

\subsection{Root Cause}

Prior work~\cite{chakraborty2023randomized, saileshwar2021mirage, purnal2019advanced} has established that pseudo-fully associative design rationale like \mirage\ is more resilient wrt. contention-based attacks compared to other contemporary designs like \scatter, \ceasers, etc. Our analysis, on the other extreme, shows that the same psuedo-fully associative design principle makes \mirage\ \textit{more} vulnerable to occupancy attacks than other variants. We now analyze the root cause of this observation and explore further this amplification of cache occupancy attacks on \mirage.

The rationale behind pseudo-fully associative design principle is to balance between security of fully associative design and efficiency of set-associative design. \mirage's design is to decouple the LLC into \emph{set-associative tag-store} and \emph{fully associative data store} and to provide extra invalid tags in the tag store. \mirage\ uses a random eviction policy for all entries in the data store, which is termed as \emph{global eviction}. Therefore, any valid data entry from the data store can get evicted. This is in contrast to fully associative caches, where valid entries can be evicted only when all invalid and free blocks have been exhausted. Likewise, this is also different from set-associative caches, where valid entries are evicted only when there is no invalid \textit{way} for the concerned \textit{set}.

Also note that such global eviction happens with very high probability for each new cache-line install, since the chances of getting set associative eviction is negligible~\cite{saileshwar2021mirage}. In other words, any incoming address is accompanied with a random eviction from the data store, which is again followed by a reverse-pointer (RPTR)-based invalidation of the corresponding entry in the tag store. As the overall occupancy of the LLC for a process grows, the chances that valid addresses from other processes already installed in the cache will get evicted also increase. For other caches, including traditional and randomized designs (e.g. \scatter, \ceasers) that are designed based on the set-associative cache primitive, eviction occurs only when all the available ways in a particular set get occupied with valid entries.

\begin{tcolorbox}[
    colback=lightgray,   
    colframe=customblack,       
    arc=2mm,             
    boxrule=1pt,          
    left=1pt,   
    right=1pt,  
    top=1pt,    
    bottom=1pt, 
]

\textbf{Takeaway: } \mirage, due to its policy of compulsory eviction from the data store for every new address installation, ensures a \textit{higher} probability of evicting a valid cache-line than any other design under consideration. As such, it becomes more vulnerable to cache occupancy attacks than other designs and even the traditional set-associative cache.
\end{tcolorbox}

Such \textit{higher} probability of evicting a valid cache-line can be further exploited to improve the bandwidth of the covert channel even with low LLC occupancy like $10\%$. We defer to Appendix~\ref{sec:template} for details.

\section{Case 2: Process fingerprinting}
\label{sec:process_fingerprinting}

Our results on covert channel experiments make it clear that \textit{all} cache design variants are vulnerable to cache occupancy attacks, with \mirage\ being the most vulnerable. In this section, we extend this result to \emph{template-based fingerprinting of processes.} The attack rationale is that different processes exhibit different number of \textbf{load}s/\textbf{store}s during their lifetime. As demonstrated in Sec.~\ref{sec:covert_channel}, such differences in the number of memory accesses exert a statistically distinguishable influence in the cache hit/miss ratio in the LLC occupancy of an attacker, and leading to process fingerprinting.

We use gem5 for this evaluation. Our gem5 setup is similar to the setup followed for benchmarking strategy \textbf{B2}:

\noindent- \textbf{Platform}: gem5\\
- \textbf{Workload}: \textcolor{black}{Spurious occupancy} followed by \spec\\
- \textbf{L1 size}: L1d : $512$ KB, L1i : $32$ KB\\
- \textbf{LLC size}: $16$ MB\\
- \textbf{LLC replacement policy} (wherever applicable): random replacement \\
- \textbf{Skews} (wherever applicable): $2$\\
- \textbf{Encryption latency}: $3$\\
- \textbf{CPU model}: TimingSimpleCPU\\

\smallskip

\begin{figure}[t]
    \centering
    \includegraphics[scale=0.285]{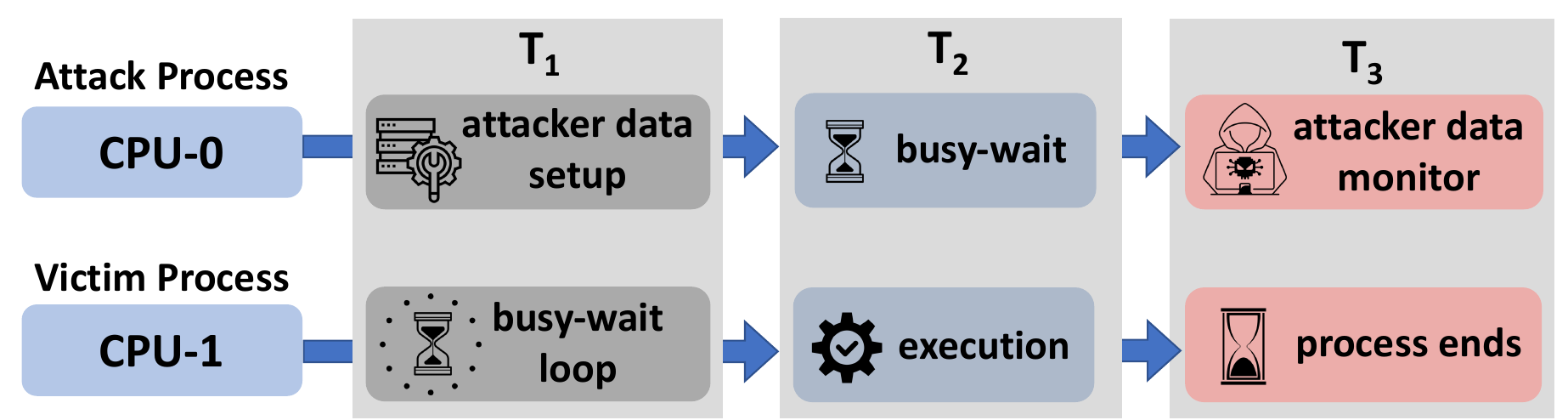}
    \caption{Schematic of experimental setup used for fingerprinting attacks on \spec\ workloads. Phases $T_{1}$, $T_{2}$, and $T_{3}$ ensure synchronization between (otherwise independent) victim and attacker.}
    \label{fig:attack_model}
\end{figure}

In order to launch the attack, we create two process binaries: \textit{attacker process} and \textit{victim workload}, as depicted in Fig.~\ref{fig:attack_model}. The simulation is run with $2$ CPUs, with each process bound to one CPU. This ensures sharing of the LLC and a realistic attack setting. Moreover, in order to ensure synchronization, we use busy-wait loops enabled by use of \texttt{rdtsc} instructions. Note that \textit{use of busy-wait loops in the victim workload is not a violation of our threat model}. In a realistic system, the micro-architectural state of the LLC would be correctly initialized, allowing the attack to launch instantaneously (i.e. not requiring any warm-up phase, and thereby no busy-wait in the victim). In gem5, however, the synchronization between the attacker and the victim is not possible, as the simulation assigns and simulates both processes on parallel cores from the very first tick of the simulation. Hence, we use \textit{busy-wait} on victim process to force synchronization. On a realistic system, the assumption that the \textit{attacker can trigger the victim} is made to ensure synchronization. This assumption is in line with other works in literature on similar attack vectors.

\begin{figure}[!t]
    \centering
    \includegraphics[scale=0.4]{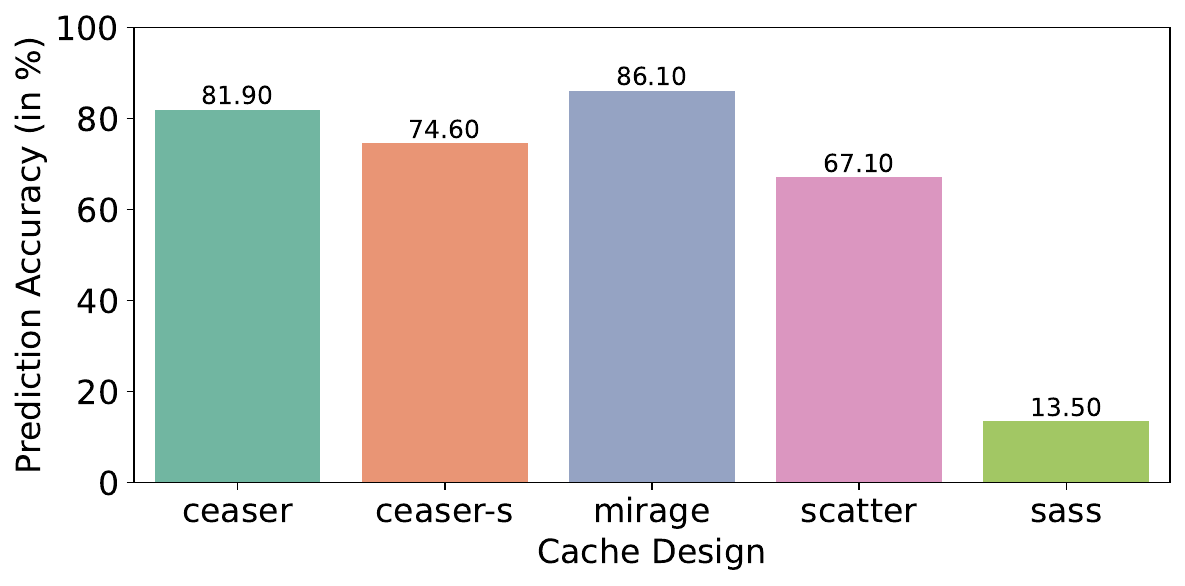}
    
    \caption{Comparative analysis of fingerprinting accuracy across different cache designs across $500$ predictions of randomly sampled workloads from \spec.}
    \label{fig:fingerprinting_accuracy}
\end{figure}


The fingerprinting attack proceeds in two phases: offline and online phase. In the offline phase, the \textit{attacker process} on CPU-$0$ creates templates for different workloads by observing LLC misses (observed through gem5 metric \texttt{system.l2.overall\_misses::.cpu0.data}) for $15\%$ LLC occupancy. In the online phase, it uses these templates to predict the workload being executed on CPU-$1$. Fig.~\ref{fig:fingerprinting_accuracy} captures the relative accuracies of different cache designs for $500$ predictions of workloads randomly sampled from \spec.



\begin{tcolorbox}[
    colback=lightgray,   
    colframe=customblack,       
    arc=2mm,             
    boxrule=1pt,          
    left=1pt,   
    right=1pt,  
    top=1pt,    
    bottom=1pt, 
]

\textbf{Takeaway: } With respect to process fingerprinting attack through cache occupancy, only \sass~exhibits a remarkably less prediction accuracy, implying no statistically significant adversarial advantage. All other cache designs exhibit vulnerability, with \mirage\ empirically being highly vulnerable, and \scatter\ being (comparatively) less vulnerable.
\end{tcolorbox}

We attribute the resilience of \sass~to our fingerprinting attack to its design decision of security-domain isolated subseting of LLC. Since the amount of \textit{overlap} between the two processes in Fig.~\ref{fig:attack_model} is bounded, there is limited leakage of victim workload through attacker's cache occupancy channel.





\section{Case 3: Key recovery on AES}~\label{sec:aes_attack}

We now detail our experiments wrt. using cache occupancy side-channel to recover AES-$128$ bit key. While prior work~\cite{ramkrishnan2024non,genkin2022cachefx} has demonstrated the vulnerability of randomized caches to occupancy attacks by distinguishing between two keys, but they do not report key recovery which would be expected for a practical attack.

\subsection{Attack Description}
\label{sec:aes_attack_description}

We use gem5 for this evaluation. Our gem5 setup is similar to the setup followed for benchmarking strategy \textbf{B2}) in Sec.~\ref{sec:benchmarking_spurious_occupancy}:

\noindent- \textbf{Platform}: gem5\\
- \textbf{Workload}: \textcolor{black}{Spurious occupancy} followed by \textbf{AES victim}\\
- \textbf{L1 size}: L1d : $512$ KB, L1i : $32$ KB\\
- \textbf{LLC size}: $16$ MB\\
- \textbf{LLC replacement policy} (wherever applicable): random replacement \\
- \textbf{Skews} (wherever applicable): $2$\\
- \textbf{Encryption latency}: $3$\\
- \textbf{CPU model}: TimingSimpleCPU\\

\smallskip

Our \textbf{AES victim}, like~\cite{ramkrishnan2024non,genkin2022cachefx}, is an OpenSSL styled T-Table implementation. It accepts a randomly generated plaintext, and for a fixed secret key, performs a \textit{single} encryption and exits. Our \textbf{attacker} is a novel variant of cache timing attacks on AES T-Table implementations~\cite{aly2013attacking}, where instead of timing victim AES execution, we rather time the accesses to an attacker's LLC occupancy. Algo.~\ref{algo:aes_attack} (in Appendix) summarizes the sequence of operations performed by the attacker, which we now explain:

\begin{enumerate} \itemsep0em
    \item The attacker first lets the AES victim setup its secret key, and precompute T-Tables.

    \item For the same reasons as detailed in Sec.~\ref{sec:mirage_assumptions}, before the experiment begins, the attacker fills the LLC with spurious occupancy. These memory accesses are \textit{never} re-accessed again during the course of the actual attack. This step also ensures that the AES victim T-Tables are reliably flushed from the LLC.

    \item Given a fixed occupancy $X\%$, the attacker \texttt{malloc}s about $\frac{16\times X}{100}$ MB of memory, and repeatedly accesses it to ensure occupancy of complete L1d and $X\%$ of LLC, since LLC is inclusive cache\footnote{Most cache designs (including the ones we evaluate here) consider an inclusive cache-hierarchy.}.

    \item Given a randomly generated plaintext $P$, the attacker then lets AES victim run a \textit{single} encryption of $P$, and obtains the ciphertext $C$.

    \item Finally, the attacker uses \texttt{rdtsc} to time access to its previously allocated $\frac{16\times X}{100}$ MB of memory. Call it $T$.
\end{enumerate}

A single observation point for the attacker is the tuple $(P, C, T)$: the time $T$ taken to access its \textit{own} cache occupancy \textit{after} the AES victim operates upon plaintext $P$ to generate $C$, for an \textbf{unknown key} $K$\footnote{Without loss of generality, we target recovery of last round AES key. Keys of other rounds can be recovered through the AES key schedule once last round key is recovered.}. The adversary repeats the aforementioned steps and obtains an observation trace $O = \{(P_{1}, C_{1}, T_{1}), (P_{2}, C_{2}, T_{2}), (P_{3}, C_{3}, T_{3}), \cdots, (P_{\mid O \mid}, C_{\mid O \mid}, T_{\mid O \mid} ) \}$. Thereafter, the attacker \textit{repeats} the same steps as above, but rather than invoking the AES victim, it \textit{simulates} AES T-Table encryption for a \textbf{known key} $K^{*}$. It builds another observation trace $O^{*} = \{(P^{*}_{1}, C^{*}_{1}, T^{*}_{1}), (P^{*}_{2}, C^{*}_{2}, T^{*}_{2}), (P^{*}_{3}, C^{*}_{3}, T^{*}_{3}), \cdots, (P^{*}_{\mid O^{*} \mid}, C^{*}_{\mid O^{*} \mid}, T^{*}_{\mid O^{*} \mid} ) \}$. Recovery of the \textbf{unknown key} $K^{*}$ then proceeds as follows:

\begin{enumerate} \itemsep0em
    \item For each $(P^{*}_{i}, C^{*}_{i}, T^{*}_{i}) \in O^{*}$, the attacker first computes $X_{i}$ = \texttt{InvSBox}($C^{*}_{i} \oplus K^{*}$) to derive the output of the \texttt{SBox} operation in the last round of AES. The adversary thus has $X = \{X_{1}, X_{2}, \cdots, X_{\mid O^{*} \mid}\}$.

    \item For each key byte $\{b_{j} \in K^{*}: 0 \le j \le 15\}$, the attacker creates a dictionary $\mathcal{T}^{*}_{j}$, where labels range from $\{0, 1, 2, \cdots, 255\}$. Against each label $l \in \{0, 1, 2, \cdots, 255\}$, $\mathcal{T}^{*}_{j}[l]$ stores the \textit{mean} of all timing values $\{T^{*}_{(1)}, T^{*}_{(2)}, T^{*}_{(3)}, \cdots\}$ for which the corresponding entry in $X$ equals $l$. At the end of this phase, the attacker has $16$ dictionaries $\{\mathcal{T}^{*}_{0}, \mathcal{T}^{*}_{1}, \mathcal{T}^{*}_{2}, \mathcal{T}^{*}_{3}, \cdots, \mathcal{T}^{*}_{15}\}$.

    \item For the observation trace $O$ for the \textbf{unknown key} $K$, the attacker follows a similar procedure as in \textcircled{2}. The only difference is that since the key is \textbf{unknown}, the attacker makes \textit{guesses} for \textit{each} of the key byte. This leads to $256$ templates for \textit{each} key byte, implying the total number of templates created by the adversary is $(256 \times 16)$. Call this template set: $\{\mathcal{T}_{j}^{i}: 0 \le i \le 15, \, 0 \le j \le 255$\}.

    \item Finally, for each key byte index $i \in \{0, 1, 2, \cdots, 15\}$, the attacker correlates $\{\mathcal{T}^{i}_{0}, \mathcal{T}^{i}_{1}, \mathcal{T}^{i}_{2}, \mathcal{T}^{i}_{3}, \mathcal{T}^{i}_{4}, \cdots, \mathcal{T}^{i}_{255}\}$ with $\mathcal{T}^{*}_{i}$. At this phase, the attacker sorts the key guesses based on the achieved correlation, and outputs the \textit{ranks} of the correct key bytes as $R = \{R_{0}, R_{1}, R_{2}, \cdots, R_{15}\}$.
\end{enumerate}

\begin{algorithm}[!h]
\caption{Algorithmic sketch of the AES attack. \texttt{AESEncrypt} and \texttt{ComputeGuessingEntropy} are generic procedures, and their descriptions are omitted for brevity.}
\label{algo:aes_attack}
\begin{algorithmic}[1]

\Procedure{createKnownKeyProfile}{}
   \State Initialize dictionaries $\{\mathcal{T}_{i}^{*}: 0 \le i \le 15\}$
   \For{\texttt{observation} \textbf{in} $\{1, 2, \cdots, \mid O^{*}\mid\}$}
      \State $P \leftarrow$ randomly generated plaintext
      \State /* \textbf{Adversary State Preparation} */
      \State malloc \texttt{data} of size $X\%$ of LLC
      \For{\texttt{index} \textbf{in} $\{1, 2, \cdots, \mid data\mid\}$}
        \State load \texttt{data[index]}
    \EndFor
    \smallskip
    \State /* \textbf{AES Encryption with known key} $K^{*}$ */    
    \State $C$ = \texttt{AESEncrypt}($P$, $K^{*}$)
    
    \smallskip
    \State /*\textbf{ Adversarial Post-Processing}/*
    \State \texttt{start\_time} $\leftarrow$ \texttt{rdtsc()}
    \For{\texttt{index} \textbf{in} $\{1, 2, \cdots, \mid     data\mid\}$}
        \State load \texttt{data[index]}
    \EndFor
    \State $T$ $\leftarrow$ \texttt{rdtsc()} - \texttt{start\_time}
    \State \texttt{state} = \texttt{InvSBox(}$C \oplus K^{*}$)

    \State \texttt{dict\_index} $\leftarrow$ 0
    \For{\texttt{byte} \textbf{in} \texttt{state}}
    \State $\mathcal{T}^{*}_{\mathtt{dict\_index}}[\mathtt{byte}] = \mathcal{T}^{*}_{\mathtt{dict\_index}}[\mathtt{byte}] \, \,\cup \,\, T$
    \State \texttt{dict\_index} $\leftarrow$ \texttt{dict\_index} + $1$
    \EndFor
   \EndFor
   \State \textbf{return} $\{\mathcal{T}^{*}_{i}\}$
\EndProcedure

\Procedure{createUnknownKeyProfile}{}
   \State Initialize dictionaries $\{\mathcal{T}_{j}^{i}: 0 \le i \le 15, 0 \le j \le 255\}$
   \For{\texttt{observation} \textbf{in} $\{1, 2, \cdots, \mid O\mid\}$}
      \State $P \leftarrow$ randomly generated plaintext
      \State /* \textbf{Adversary State Preparation} */
      \State malloc \texttt{data} of size $X\%$ of LLC
      \For{\texttt{index} \textbf{in} $\{1, 2, \cdots, \mid data\mid\}$}
        \State load \texttt{data[index]}
    \EndFor
    \smallskip
    \State /* \textbf{AES Encryption with unknown key} $K$ */    
    \State $C$ = \texttt{AESEncrypt}($P$, $K$)
    
    \smallskip
    \State /*\textbf{ Adversarial Post-Processing}/*
    \State \texttt{start\_time} $\leftarrow$ \texttt{rdtsc()}
    \For{\texttt{index} \textbf{in} $\{1, 2, \cdots, \mid     data\mid\}$}
        \State load \texttt{data[index]}
    \EndFor
    \State $T$ $\leftarrow$ \texttt{rdtsc()} - \texttt{start\_time}

    \For{\texttt{guess} \textbf{in} $\{0, 1, \cdots, 255\}$}
    
    \State \texttt{state} = \texttt{InvSBox(}$C \oplus$ \texttt{guess})

    \State \texttt{dict\_index} $\leftarrow$ 0
    \For{\texttt{byte} \textbf{in} \texttt{state}}
    \State $\mathcal{T}^{\mathtt{dict\_index}}_{guess}[\mathtt{byte}] = \mathcal{T}^{\mathtt{dict\_index}}_{guess}[\mathtt{byte}] \, \,\cup \,\, T$
    \State \texttt{dict\_index} $\leftarrow$ \texttt{dict\_index} + $1$
    \EndFor
    \EndFor
   \EndFor
   \State \textbf{return} $\{\mathcal{T}_{j}^{i}: 0 \le i \le 15, 0 \le j \le 255\}$
\EndProcedure

\Procedure{attack}{}
    \State \texttt{InitializeAESTables()}
    \State malloc \texttt{data} of LLC size
    \For{\texttt{index} \textbf{in} $\{1, 2, \cdots, \mid data\mid\}$}
        \State load \texttt{data[index]}
    \EndFor
    \State $\mathcal{T}^{*}_{i}$ $\leftarrow$ \texttt{createKnownKeyProfile()}
    \State $\mathcal{T}_{j}^{i}$ $\leftarrow$ \texttt{createUnknownKeyProfile}()
    \State \texttt{ComputeGuessingEntropy}($\mathcal{T}^{*}_{i}$, $\mathcal{T}_{j}^{i}$)
\EndProcedure
\end{algorithmic}
\end{algorithm}

A successful key recovery is achieved when all key byte ranks are $1$. However, a better indicator of whether a side-channel leaks is the guessing entropy~\cite{liu2022frequency}:

$$\text{GE} = \Sigma_{i=0}^{15} \, log_{2}(R_{i})$$

Lower GE implies the key bytes are lower ranked post correlation. In practice, achieving a GE lesser than a certain pre-defined threshold is sufficient to declare an implementation vulnerable to successful key recovery with reasonable computational complexity~\cite{liu2022frequency,standaert2009unified}. Usually a GE of about $32$ is considered to be the threshold~\cite{liu2022frequency}; it implies the correct key bytes (on average) are among the top $4$ guessed key bytes by the adversary. A side-channel achieving a GE of $32$ can reliably converge to GE = $0$ by increasing the number of observations~\cite{liu2022frequency}. Furthermore, even at GE of about $32$, the brute force complexity of the key-space has reduced from $2^{128}$ to $2^{32}$, which is tractable on modern systems (refer Sec.~\ref{sec:full_key_recovery}).

We now define parameters for estimating \textit{adversarial complexity} for mounting this attack. This complexity can be influenced by adjusting either of the following parameters:

\begin{itemize}
    \item $\%$\textbf{ of LLC that the adversary is allowed to occupy}: A higher occupancy makes it easier to observe activity of the victim program (cf. Sec.~\ref{sec:process_fingerprinting}), but is easier to detect by defence mechanisms.
    
    \item \textbf{Number of observations that the adversary is allowed to measure}: Note that the success of the attack higher depends upon $\mid O \mid$ and $\mid O^{*} \mid$. Higher $\mid O \mid$ and $\mid O^{*} \mid$ imply better chances at convergence, but increases the required time to attack.
\end{itemize}

We treat both these considerations independently now.

\subsection{Key Recovery Under Varying Occupancy}
\label{sec:key_recovery_varying_occupancy}

We first validate our attack against a varying number of LLC occupancy rates. Since our objective is to just choose an optimal occupancy rate against which we will evaluate \textit{all} designs, we only consider three designs: \scatter, \mirage, \sass. We choose \scatter~as it is the most resilient design among all the set-associative based randomized designs we consider in this work. Likewise, we consider \mirage~as it allows us to also compare the pseudo-fully associative design rationale against set-associative based designs wrt. our attack. Finally, we consider \sass~in order to validate the extent of its resilience to occupancy attacks.

\begin{figure}[!t]
    \centering
    \includegraphics[scale=0.47]{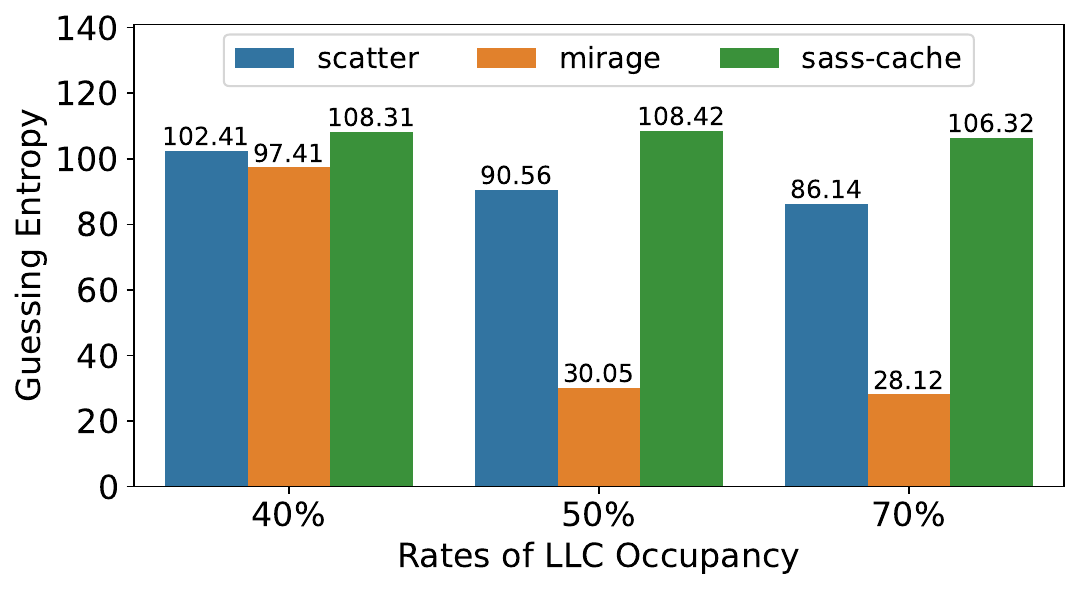}
    
    \caption{Key Recovery experiment across three occupancy rates: $40\%$, $50\%$, $70\%$ for fixed $12000$ observations.}
    \label{fig:different_occupancy_aes}
\end{figure}

Fig.~\ref{fig:different_occupancy_aes} captures our findings. First, we note that the guessing entropy of \sass~shows limited improvement with increasing attacker occupancy. This can be attributed to its security-domain isolation of LLC occupancy. Next, \scatter\ does show improvement as the cache occupancy increases, but the Guessing Entropy is not sufficiently low to perform any key recovery. Finally, \mirage's Guessing Entropy falls as the attacker occupancy increases, eventually reaching the chosen threshold of $32$. Note however that we were able to observe leakage only for occupancy rates $> 50\%$\footnote{We need a higher LLC occupancy than the attacks in Sec.~\ref{sec:covert_channel} and Sec.~\ref{sec:process_fingerprinting} needed because victim AES has relatively low memory footprint than the workloads considered in Sec.~\ref{sec:covert_channel} and Sec.~\ref{sec:process_fingerprinting}.}.

\subsection{Key Recovery Under Varying Number of Observations}

We now fix the attacker LLC occupancy at a conservative $50\%$, and evaluate the \textit{trends} in Guessing Entropy against \textit{all} designs considered in this work: \ceasers, \scatter, \mirage, and \sass. We vary the number of observations ($\mid O \mid$ and $\mid O^{*} \mid$) the adversary is allowed to make from $100$ to $20000$\footnote{We scheduled about $350$ single-threaded gem5 simulations in parallel, allowing us to collect observations at a rate of $500$ observations per hour. Such inhibitory rate is because we work with gem5; on an actual LLC, the rate of collection of observations would be much higher.} and repeat the attack on all cache designs.

\begin{figure}[!t]
    \centering
    \includegraphics[scale=0.38]{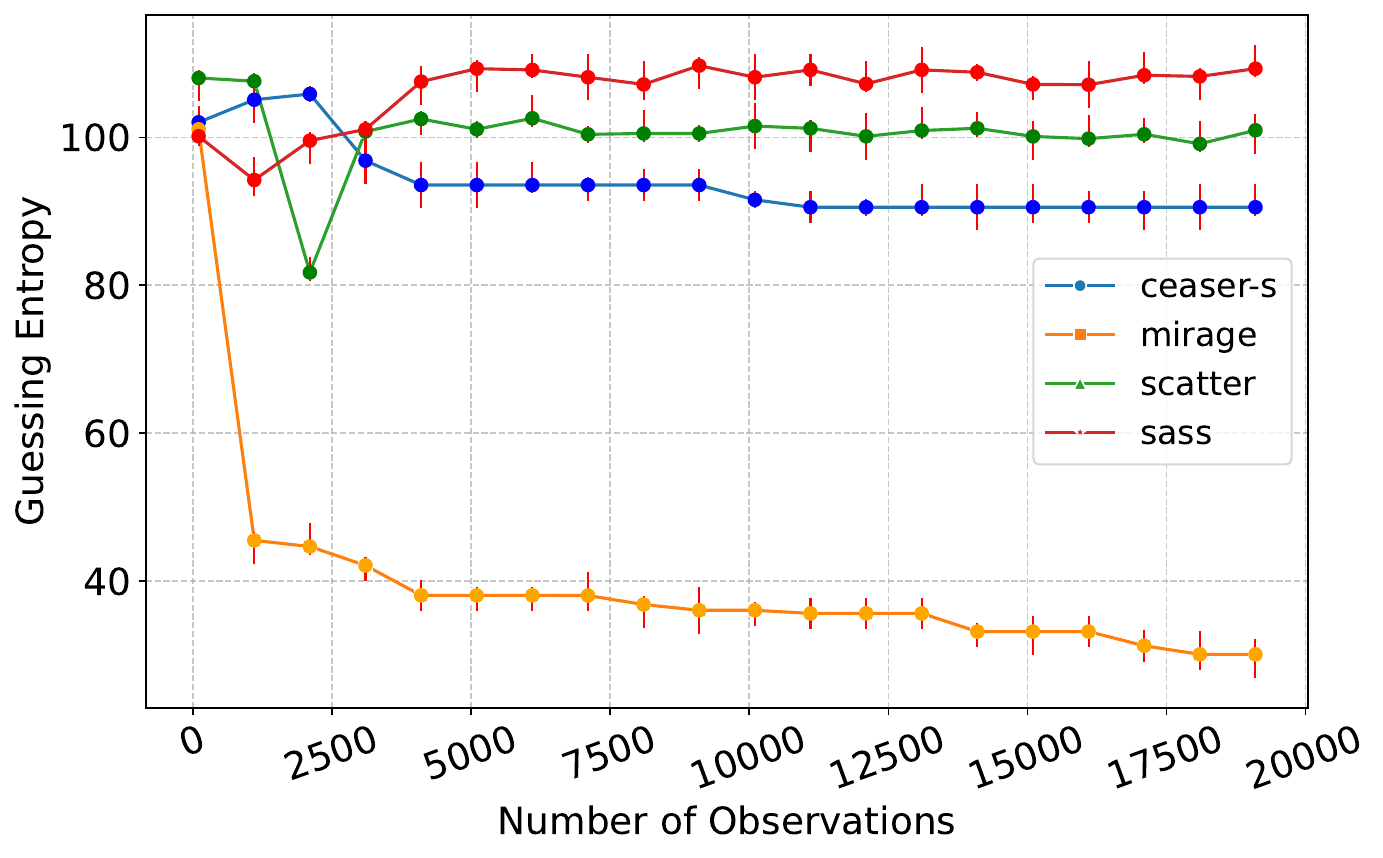}
    
    \caption{Key Recovery experiment across a fixed occupancy rate: $50\%$, and for varying number of observations.}
    \label{fig:fixed_occupancy_aes}
\end{figure}

Our results are summarized in Fig.~\ref{fig:fixed_occupancy_aes}. First, \sass~never improves even upon increasing the number of observations. Then between \ceasers~and \scatter, both improve from their initial value, depicting \textit{some} leakage, but not enough to mount any key recovery attack. Finally, \mirage~improves significantly, and eventually crosses the GE threshold of $32$, implying all secret key bytes are on average ranked within the top $4$ key guesses by the adversary.

\begin{tcolorbox}[
    colback=lightgray,   
    colframe=customblack,       
    arc=2mm,             
    boxrule=1pt,          
    left=1pt,   
    right=1pt,  
    top=1pt,    
    bottom=1pt, 
]

\textbf{Takeaway: } Comparing between designs based on set-associative caches and designs based on pseudo-fully associative caches, our evaluations demonstrate that the former is more resilient to our AES key recovery attack than the latter. This takeaway is consistent with the takeaways in Sec.~\ref{sec:covert_channel}  and Sec.~\ref{sec:process_fingerprinting}.

Between the different designs within the class of set-associative based randomized caches, \sass~is the most secure (against all evaluated levels of occupancy), validating its design rationale as the best possible choice to protect against \textit{both} contention and occupancy attacks.
\end{tcolorbox}

\subsection{Full Key Recovery}
\label{sec:full_key_recovery}

In \mirage\, each key byte is on average among the top $4$ ranks, making the brute-force complexity of AES for \mirage\ to be essentially $2^{32}$. We can simply brute force the secret key at this point. We deploy four Intel(R) Xeon(R) Gold $6226R$ CPU servers (running Ubuntu 20.04.6 LTS, $32$ GB RAM), with a cumulative logical core count of about $350$. We parallelize the brute force search across all logical cores, reducing the workload complexity of each logical core to $\frac{2^{32}}{2^{8}} \approx 2^{24}$. In about $6$ hours, we were able to leak the entire secret key.

Note from Fig.~\ref{fig:fixed_occupancy_aes} that a similar attack is infeasible for \ceasers~and \scatter, because although their attack complexity has reduced significantly from $2^{128}$ (thereby establishing leakage), it is still computationally infeasible to recover the key. However, as with increasing number of observations, the GE shows a downward trend, it is an interesting future problem to estimate the number of adversarial observations required to achieve the same Guessing Entropy in \ceasers~and \scatter\ as we observe with \mirage.  

\section{Discussion and Future Directions}
\label{sec:discussion}

Considerations of cache occupancy in designs of randomized caches have largely been overlooked in contemporary literature. However, the security evaluations we perform in this work (especially related to occupancy-based full key recovery attack on AES) reliably reinforce the need to consider \textit{both} occupancy attacks and contention-based attacks in randomized cache designs. 

Our results have a few recurring and novel takeaways. First, \textcircled{1} some set-associative randomized caches (like \scatter~and \ceasers) can \textit{outperform} baseline set-associative caches when paired with the correct replacement policies. Prior works only report slowdown in their comparative analysis, since they usually don't consider replacement policies as a parameter in their evaluation. Secondly, \textcircled{2} pseudo-fully associative cache designs like \mirage\ are worse off than designs based on set-associativity wrt. resilience against occupancy attacks. Finally, \textcircled{3} \sass~resists all our attack vectors. This is expected; \sass~implements security-domain based non-overlapping partitions. As such, each security-domain has a fixed subset of LLC it can occupy for operation; this design decision gives good security against \textit{both} contention-based and occupancy-based attacks. However, our performance evaluation on realistic LLC configurations also establishes that this very design decision is prohibitive in terms of \sass's performance across all considered replacement policies.

These findings thereby establish an interesting open problem: Design of a randomized cache of comparable efficiency with modern set-associative LLCs, while still resisting \textit{both} contention-based and occupancy-based attacks. We believe the design rationale of dynamic partitioning of the LLC may provide solutions to the questions we raise in this work. Randomized caches designed around dynamically changing partitions would be capable of restricting LLC occupancy for critical workloads (thereby providing security), while also generously allowing LLC occupancy for non-critical workloads (thereby providing better performance in general). More research however is required to concretely establish how such a cache design shall function, how process requests for dynamic changes in partitioning will be handled, and how to implement this design in an uncomplicated manner.

\section{Conclusion}
\label{sec:conclusion}

The design motivation of state-of-the-art cache randomization schemes is majorly directed towards protecting against contention-based attacks while essentially ignoring other cache attack variants that are equally practical and powerful. In this work, we walk a different route by evaluating the relative resilience of contemporary secured cache schemes against \emph{cache occupancy attacks}. We do so in two verticals: performance and security. For our performance evaluation, we first establish that existing benchmarking strategies provide an unfair perspective on comparative performance (due to varying LLC configurations, choice of benchmarks, choice of simulation platform, implementation-specific assumptions etc). We thus propose a uniform benchmarking strategy for fair evaluation, and evaluate five cache designs: \ceaser, \ceasers, \mirage, \scatter, and \sass~across five replacement policies. For our security evaluation, we evaluate the resistance of these five cache designs against different threat models, and comparatively analyze the resilience of different cache designs against our attacks. Moreover, for the first time, we show full AES key recovery attack through a novel cache-occupancy side-channel, thereby establishing the potency of occupancy-based attacks on randomized caches.

Our results therefore highlight the need for a holistic performance and security evaluation as we demonstrate that design decisions to protect against contention-based attacks alleviate vulnerability towards cache occupancy attacks as well as slowdown in performance in general.

\section*{Acknowledgments}

We would like to thank the authors of \sass~who very generously provided us with their custom simulator for \sass~upon request, which allowed for a fair performance and security evaluation across all designs. We would also like to thank the anonymous reviewers for their constructive comments that helped in improving the overall message of our work. Nimish Mishra, Sarani Bhattacharya, and Debdeep Mukhopadhyay would additionally like to acknowledge Centre on Hardware-Security Entrepreneurship Research and Development
(C-HERD), MeitY, Govt. of India, and Information Security Education and
Awareness (ISEA), MeitY, Govt. of India, for partially funding this research.

\section*{Ethical Consideration and Open Science}
\subsubsection*{Ethical Consideration}
The authors acknowledge and uphold the importance of maintaining high ethical standards in conducting and evaluating computer security research.
This work analyses multiple secured cache designs proposed in literature over the past decade. None of the designs-under-test are part of commercially available product and thus require no vulnerability disclosure. The designs implemented using in-house simulator are based on the design descriptions from individual papers. The gem5 implementations are adapted from the respective open-source repositories associated with original papers and have been appropriately attributed, wherever applicable. 

\subsubsection*{Open Science Compliance}
As part of open science initiative, we have open-sourced all the tools, scripts, and datasets developed in this work and needed for meaningful reproduction of the results shown in the paper. The permanent link for the source codes used in this paper can be found at: 
\begin{center}
    \url{https://doi.org/10.5281/zenodo.14737392}
\end{center}
Note that this permanent repository includes the codes and scripts \emph{as-is}, as used by the authors for developing the paper. A more detailed version of the artifact will be made available later, as `Artifact Appendix'. We now elucidate the components provided in the above-mentioned artifact repository.
\begin{enumerate}   \itemsep0em
    \item \textbf{llc\_simulator}: In-house simulator to produce results of Sec.\ref{sec:covert_channel}~(Fig.~\ref{thr_ref_diag}) related to covert channels. Each sub-directory within this directory corresponds to a specific randomized cache design. The results can be produced by running \texttt{python3 main.py} in each sub-directory, followed by executing the \texttt{plot.py} script.
    \item \textbf{scripts}: contains the scripts for producing plots for the following:
        \begin{enumerate}   \itemsep0em
            \item \textbf{perf\_eval}: scripts to produce benchmarking results from Sec.~\ref{sec:perf}~(Fig.~\ref{fig:random_rp}).
            \item \textbf{fingerprinting}: \texttt{accuracy.py} produces results from Sec.~\ref{sec:process_fingerprinting}~(Fig.~\ref{fig:fingerprinting_accuracy}).
            \item \textbf{aes}: scripts to produce Fig.~\ref{fig:different_occupancy_aes} and  Fig.~\ref{fig:fixed_occupancy_aes} from Sec.~\ref{sec:aes_attack}.
        \end{enumerate}
    \item \textbf{ceaser}: gem5 implementation of \ceaser.
    \item \textbf{ceaser-s}: gem5 implementation of \ceasers.
    \item \textbf{mirage}: gem5 implementations of \mirage, \scatter\ and baseline cache.
    \item \textbf{sasscache}: gem5 implementation of \sass.
\end{enumerate}

\bibliographystyle{plain}
\bibliography{refs}

\appendix

\section{Investigating Additional Assumptions in \mirage's Implementation}
\label{sec:mirage_assumptions_experiments}

We experimentally validate the extent to which \mirage's implementation specific assumptions are valid. Our experiment aims to establish how many of these memory accesses \mirage\ fails to consider in the measurement of its performance statistics. The first statistic- \texttt{system.l2.tags.data\_accesses}- denotes the total number of accesses made by the benchmark in LLC\footnote{We consider L2 cache and LLC as analogous in line with the cache configuration in \cite{saileshwar2021mirage}.}. Likewise, the second statistic- \texttt{system.l2.overall\_hits::cpu.data}- denotes the hits in LLC cache for data accesses. Finally, the remaining two statistics- \texttt{system.l2.tags.repl\_valid\_data} and \texttt{system.l2.tags.repl\_valid\_tag}- effectively denote \\``\texttt{Scenario-C}'' (GLE) and ``\texttt{Scenario-B}'' (SAE) respectively. We note from~\cite{saileshwar2021mirage} that ``\texttt{Scenario-B}'' is statistically improbable, implying (with arbitrarily high probability), all cache misses result in a global eviction (i.e. ``\texttt{Scenario-C}''). This implies the following relation holds with high probability: ``LLC accesses = LLC hits + \texttt{Scenario-C} misses''.

We capture experimental results in Tab.~\ref{tab:gem5_scenario_c_problem}. Against each benchmark, Tab.~\ref{tab:gem5_scenario_c_problem} details the overall data accesses (reported by the gem5 statistic \texttt{system.l2.tags.data\_accesses}) as well as the number of cache hits (reported by the gem5 statistic \texttt{system.l2.overall\_hits::cpu.data}). Moreover, the number of \textit{actual misses} (computed by subtracting \texttt{system.l2.overall\_hits::cpu.data} from \texttt{system.l2.tags.data\_accesses}) are also reported. Finally, the  ``\texttt{Scenario-C}'' misses, reported by the statistic \texttt{system.l2.tags.repl\_valid\_data} is shown as `Reported Global Evictions'.
As evident, the equation ``LLC accesses = LLC hits + \texttt{Scenario-C} misses'' is not satisfied by \mirage's current gem5 implementation.

\begin{table}[!t]
      \centering
  \caption{The difference in the global evictions reported by \mirage's gem5 implementation as opposed to the ideal number of misses (computed by \textit{Overall access} - \textit{Cache hits}), against \textsf{SPEC2017} benchmarks. The subset of \spec~benchmarks considered is carefully curated considering several parameters: \textcircled{1} code size (measured in kilo lines of code), \textcircled{2} programming language (C / C++ / mixture of C and C++), and \textcircled{3} application domain.}
  \label{tab:gem5_scenario_c_problem}
  \scalebox{0.7}{
  \begin{tabular}{|l|r|r|r|r|}
    \hline
    Benchmark & Overall  & Cache hits & Reported global  & Actual misses \\
    & data accesses & & evictions & \\
    \hline
    \hline
    \textbf{998.specrand\_is} & 1749227552 & 31235387 & 0 & 1717992165 \\
    \textbf{500.perlbench\_r} & 3938592619 & 57603967 & 149011 & 3880988652 \\
    \textbf{502.gcc\_r} & 4984210067 & 65119091 & 64907 & 4919090976 \\
    \textbf{505.mcf\_r} & 4083326439 & 72091393 & 548823 & 4011235046 \\
    \textbf{508.namd\_r} & 4651478331 & 82965267 & 0 & 4568513064 \\
    \textbf{511.povray\_r} & 24833452 & 431524 & 0 & 24401928 \\
    \textbf{531.deepsjeng\_r} & 3448850191 & 34224185 & 22494607 & 3414626006 \\
    \hline
\end{tabular}}
\end{table}

\begin{figure*}[!t]
  \centering
  \begin{subfigure}{0.49\textwidth} 
    \centering
    \includegraphics[width=\textwidth]{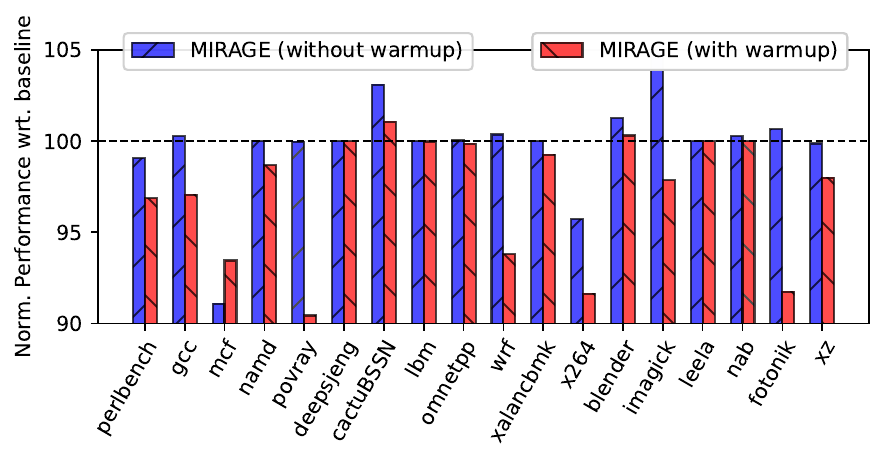} 
    \caption{\textsf{MIRAGE} vs Baseline}
    \label{fig:mirage_vs_baseline}
  \end{subfigure}
  \hfill
  \begin{subfigure}{0.49\textwidth} 
    \centering
    \includegraphics[width=\textwidth]{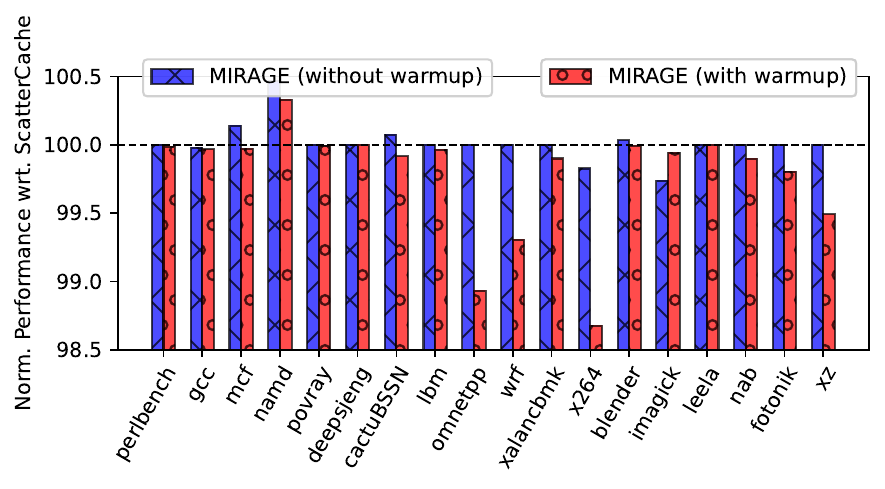}
    \caption{\textsf{MIRAGE} vs \textsf{ScatterCache}}
    \label{fig:mirage_vs_scatter}
  \end{subfigure}
  \caption{Normalized performance of \mirage\ against (a) Baseline set-associative cache and (b) \scatter\ for \spec.}
  \label{fig:mirage_perf}
\end{figure*}

To investigate the performance degradation because of this issue and provide a comparative evaluation, we run \spec\ benchmarks on gem5 implementations of \mirage, \scatter\ and baseline set-associative cache (with random replacement policy). We use the phrase \textit{cache warmup} to refer to some spurious occupancy done \textit{prior} to benchmark invocation, that helps in exhausing ``Scenario-A'' occurrences in \mirage. We note that, on average, the number of global evictions performed by \mirage, once the queue is exhausted by the \textit{cache warmup} phase, are significantly higher than reported evictions by original \mirage\ gem5 implementation. Fig.~\ref{fig:mirage_perf} shows the relative performance for \mirage\ normalized to baseline (Fig.~\ref{fig:mirage_vs_baseline}) and  \scatter\ (Fig.~\ref{fig:mirage_vs_scatter}) where the \textcolor{blue}{blue} bars denote the performance without warmup and \textcolor{red}{red} with warmup. It is clear that for most workloads, \mirage\ with warmup shows relative degradation in performance (both with respect to \scatter\ and baseline). Therefore, as Fig.~\ref{fig:mirage_perf} depicts, the \textit{original} gem5 \mirage\ implementation reports an overly optimistic performance wrt. \scatter\ and baseline cache across several \textsf{SPEC} benchmarks. Note that \mirage\ with warmup phase enabled performs worse than its original implementation as well as set associative and \scatter\ with warmup. This is because a majority of cache accesses in the original \mirage\ gem5 implementation are serviced by fully-associative design rationale (corresponding to \texttt{Scenario-A} in \mirage\ implementation) instead of being serviced by the global-eviction policy of original \mirage\ design. 

\section{Transmitting Byte-sized data in \mirage\ }
\label{sec:template}

\begin{figure}[!t]
    \centering
    \includegraphics[scale=0.56]{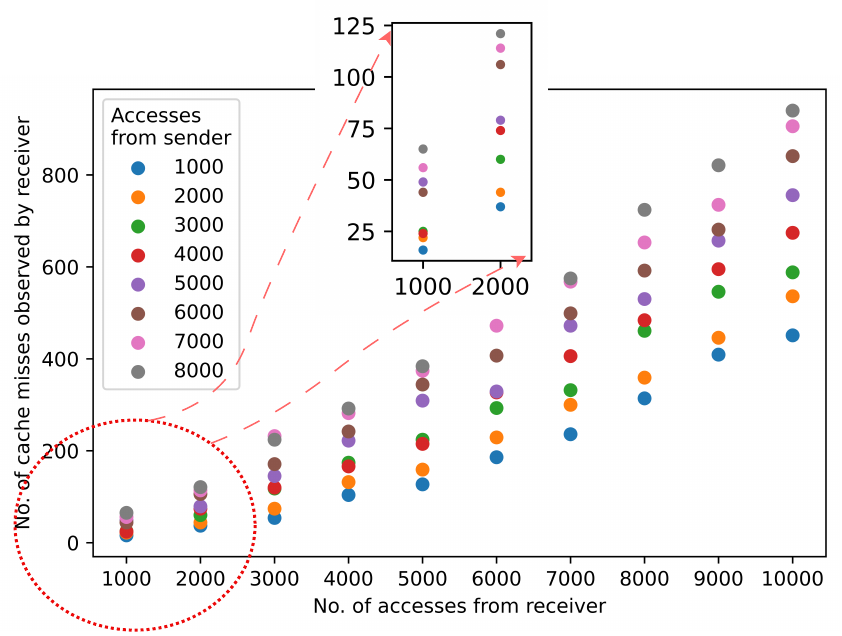}
    \caption{Covert channel in \mirage\ depicting the capability to transmit byte level information with varying number of sender and receiver accesses.}
    \label{fig:covert_channel_noiseless}
\end{figure}

\begin{figure}[!t]
    \centering
    \includegraphics[scale=0.62]{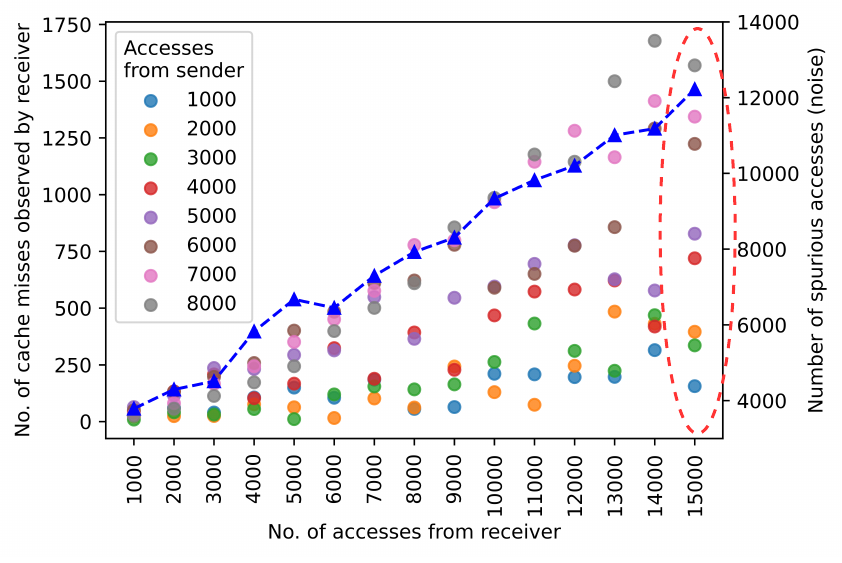}
    \caption{Covert channel in \mirage\ in noisy setting. The \textcolor{blue}{blue} line plot (\protect\filledDelta\ marker) shows the number of spurious cache accesses.}
    \label{fig:covert_channel_noisy}
\end{figure}

\begin{figure}[!t]
    \centering
    \includegraphics[scale=0.5]{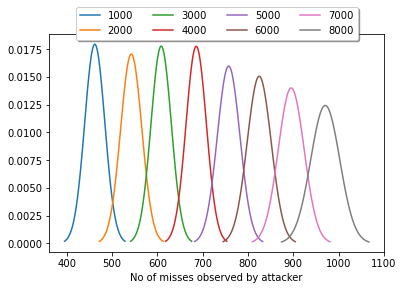}
    \caption{Templates of victim accesses with total number of accesses ranging from 1000 to 8000 and interval of 1000.}
    \label{fig:1000_template}
\end{figure}

\begin{figure}[!t]
    \centering
    \includegraphics[scale=0.35]{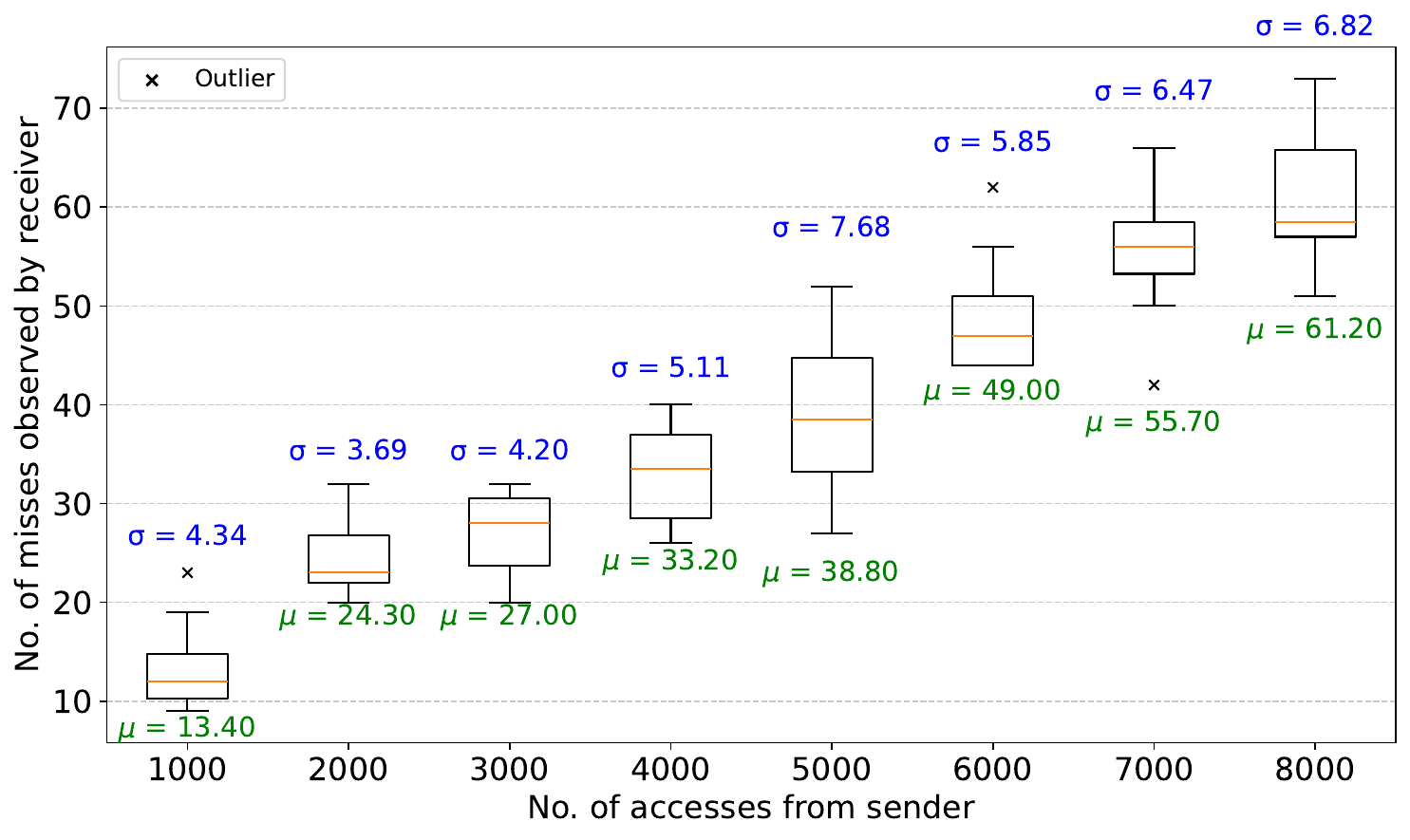}
    \caption{Covert channel statistics on \mirage\ for $1000$ accesses.}
    \label{fig:mirage_1k_access}
\end{figure}

\begin{figure}[!t]
    \centering
    \includegraphics[scale=0.35]{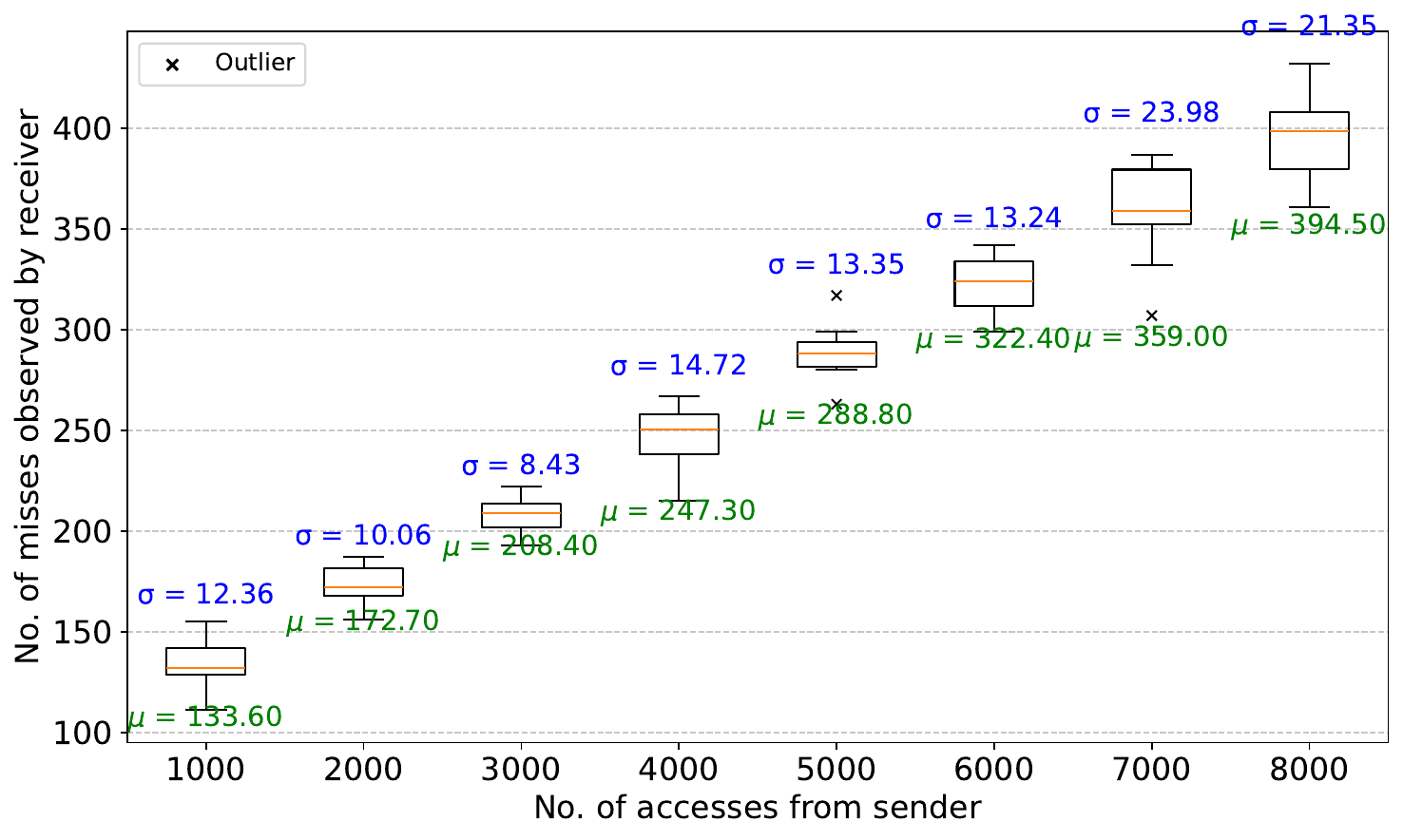}
    \caption{Covert channel statistics on \mirage\ for $5000$ accesses.}
    \label{fig:mirage_5k_access}
\end{figure}

\begin{figure}[!t]
    \centering
    \includegraphics[scale=0.35]{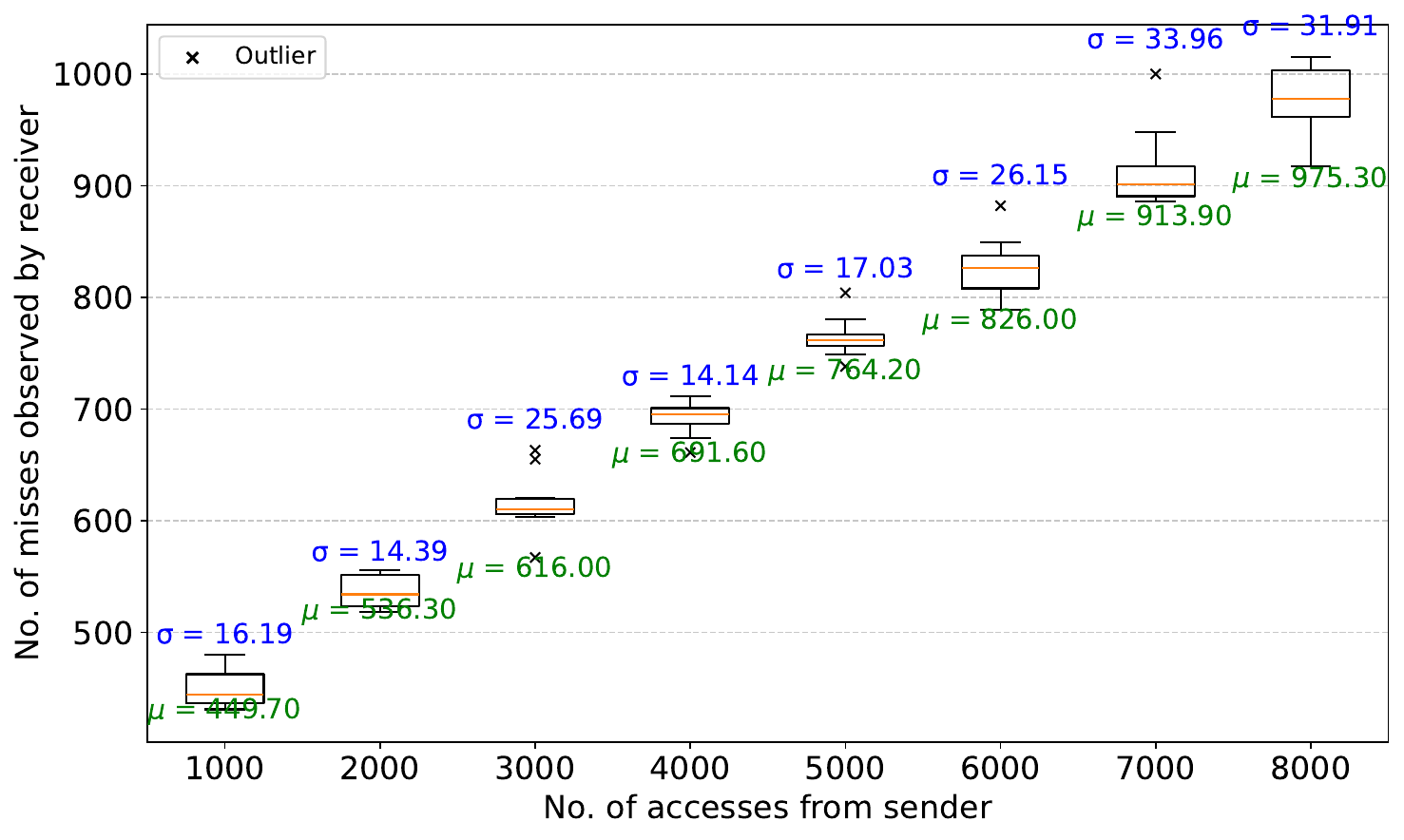}
    \caption{Covert channel statistics on \mirage\ for $10000$ accesses.}
    \label{fig:mirage_10k_access}
\end{figure}

\begin{figure}[!t]
    \centering
    \includegraphics[scale=0.35]{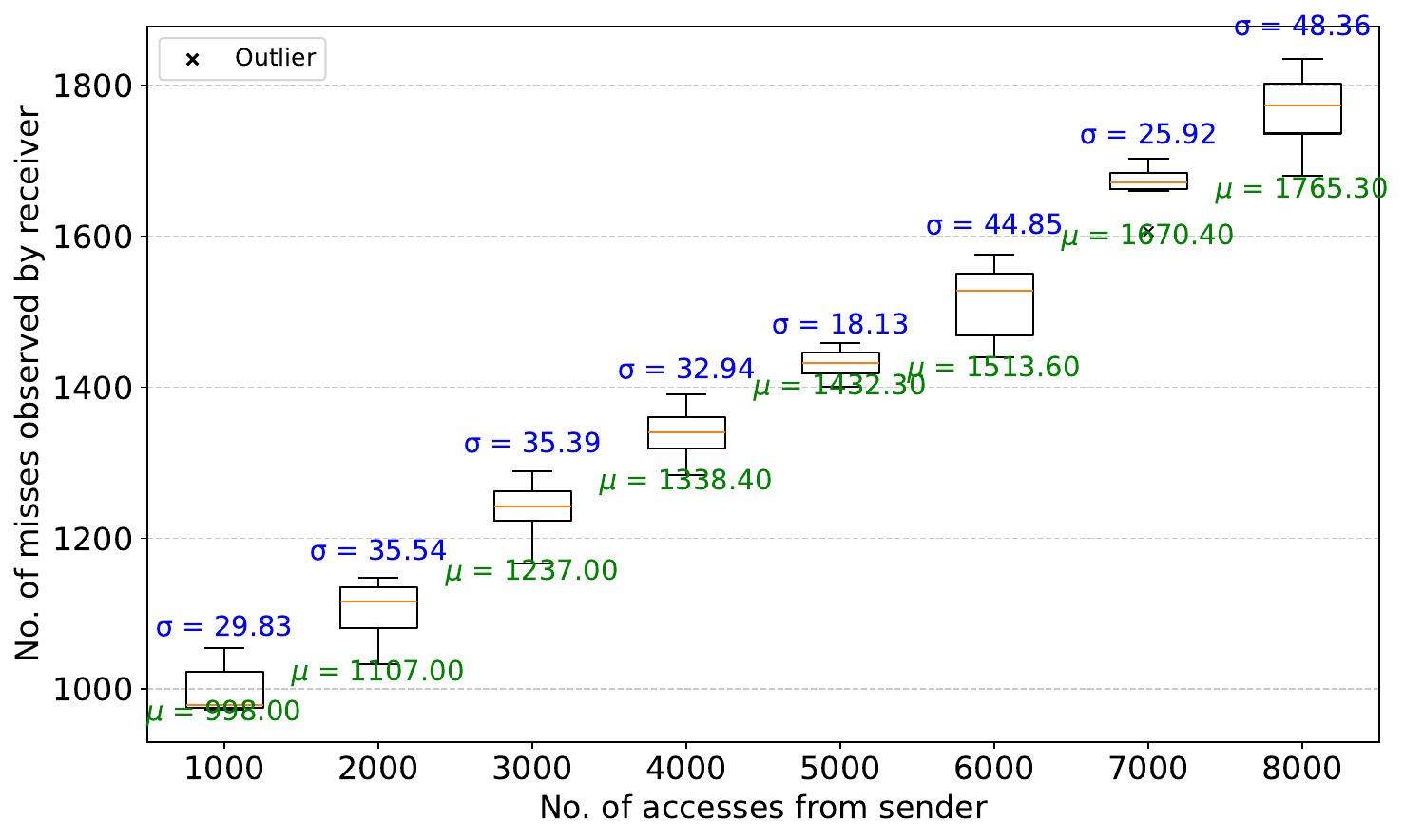}
    \caption{Covert channel statistics on \mirage\ for $15000$ accesses.}
    \label{fig:mirage_15k_access}
\end{figure}

To further evaluate the amplification of cache occupancy channel on \mirage, we extend the bit-wise covert channel into a byte-wise channel to increase the bandwidth. In this case, the receiver first prepares a \emph{pre-attack step} where it simulates different number of accesses to form templates for cache misses. 
The receiver allocates a memory space and accesses $\ell = 10,000$ addresses to install them in the cache in distinct cache lines.
The sender keeps an array of $8$ elements $\{1000, 2000, 3000, 4000, 5000, 6000, 7000, 8000 \}$, each representing a byte value as their index position in the array and the number of accesses the sender needs to make to transmit that value. For example, to transmit integer value $2$, the sender makes $3000$ accesses, whereas to transmit $7$, it makes $8000$ accesses. The rest of the process follows similar to the bit-wise communication channel. Fig.~\ref{fig:1000_template} shows the number of cache misses observed ($x$-axis) in \mirage\ by the receiver for different number of accesses made by the sender. 
While $\ell=10,000$ ensures an error-free communication channel, the channel can also perform with lesser number of receiver accesses. Fig.~\ref{fig:covert_channel_noiseless} depicts the trend of cache misses observed by the receiver with varying number of receiver accesses. As the number of accesses from the receiver increase, the reliability of the channel increases accordingly. Further, to simulate noise from other processes in real-world setting, we added random number of spurious loads. Fig.~\ref{fig:covert_channel_noisy} shows the trend of cache misses observed by the receiver. The blue lineplot shows the number of spurious accesses which we term as noise. Note that we vary the noise level in accordance with the increasing number of accesses from the receiver. Note that due to introduction of noise, the channel becomes error-free at $\ell=15,000$, instead of $10,000$ (as in Fig.~\ref{fig:covert_channel_noiseless}). The robustness of the channel against noise is due to the fact that we use the relative difference of cache misses, not their absolute values. Figs.~\ref{fig:mirage_1k_access}, \ref{fig:mirage_5k_access}, \ref{fig:mirage_10k_access}, \ref{fig:mirage_15k_access} show the statistics of the covert channel in \mirage\ for 1000, 5000, 10000, and 15000 accesses by the receiver. The boxplots show the range of cache misses observed by the receiver over $10$ iterations, with the median value depicted as an orange line inside the box. Further, the mean (\textcolor{green}{$\mu$}) and standard deviation (\textcolor{blue}{$\sigma$}) corresponding to each boxplot is annotated with green and blue, respectively. One can observe that for $10000$ and $15000$ receiver accesses (Fig. \ref{fig:mirage_10k_access}, \ref{fig:mirage_15k_access} resp.) offer almost non-overlapping statistic for the covert channel across different number of sender accesses. Whereas, for low occupancy, such as $1000$ and $5000$, receiver accesses (Fig. \ref{fig:mirage_1k_access}, \ref{fig:mirage_5k_access} resp.) show overlapping statistic, making the channel erroneous.

\begin{figure*}[!t]
    \centering
    \includegraphics[scale=0.5]{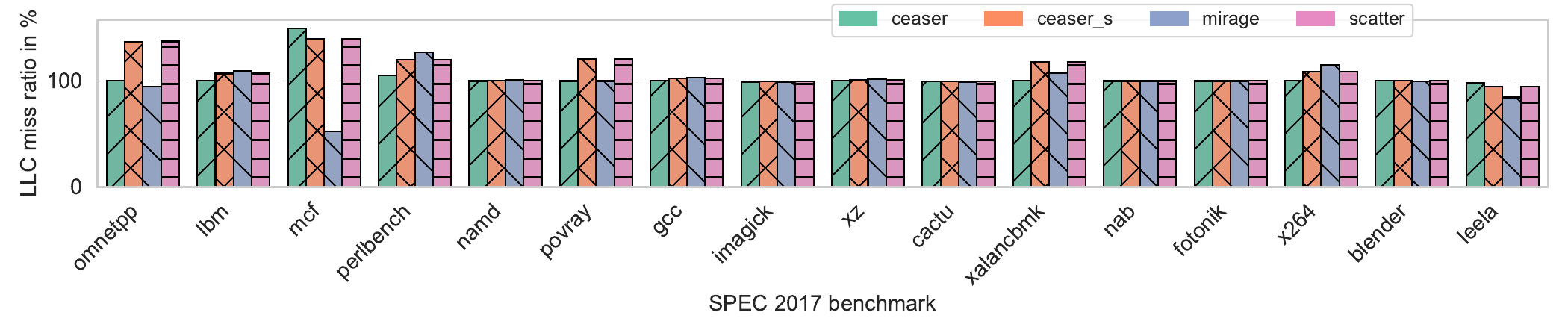}
    \caption{Performance evaluation of considered cache designs with \treePLRURP~replacement policy (normalized against baseline set-associative, and expressed as a $\%$). Performance statistics are averaged over $300$ copies of \spec~runs. Note that \sass~does not support \treePLRURP, hence is omitted for comparison here.}
    \label{fig:treePLRURP}
\end{figure*}

\begin{figure*}[!t]
    \centering
    \includegraphics[scale=0.5]{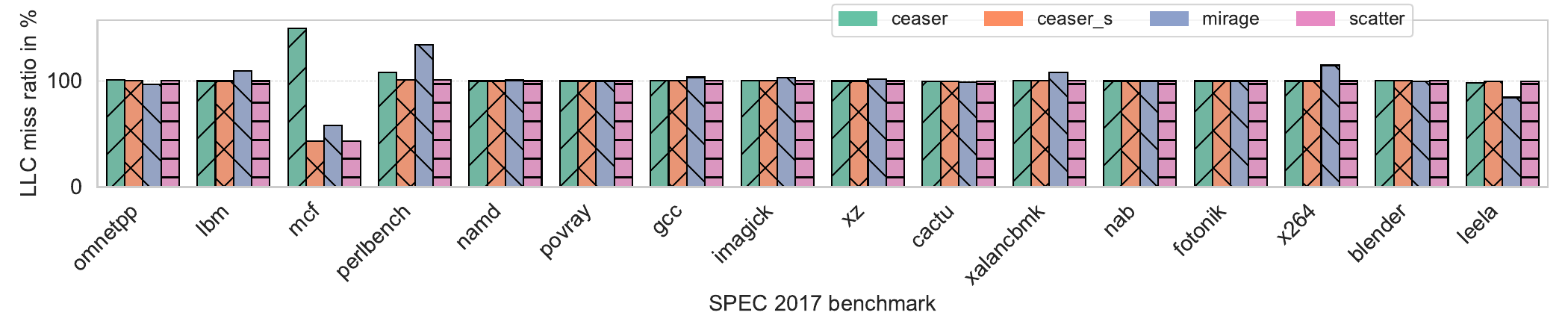}
    \caption{Performance evaluation of considered cache designs with \weightedLRURP~replacement policy (normalized against baseline set-associative, and expressed as a $\%$). Performance statistics are averaged over $300$ copies of \spec~runs. Note that \sass~does not support \weightedLRURP, hence is omitted for comparison here.}
    \label{fig:weighted_treePLRURP}
\end{figure*}

\begin{figure*}[!t]
    \centering
    \includegraphics[scale=0.5]{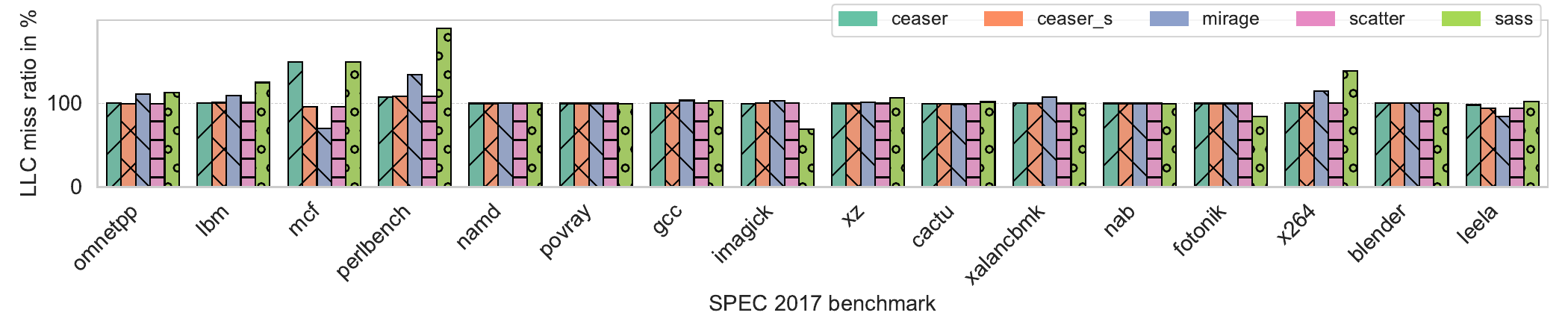}
    \caption{Performance evaluation of considered cache designs with \rriprp~replacement policy (normalized against baseline set-associative, and expressed as a $\%$). Performance statistics are averaged over $300$ copies of \spec~runs.}
    \label{fig:rriprp}
\end{figure*}

\begin{figure*}[!t]
    \centering
    \includegraphics[scale=0.5]{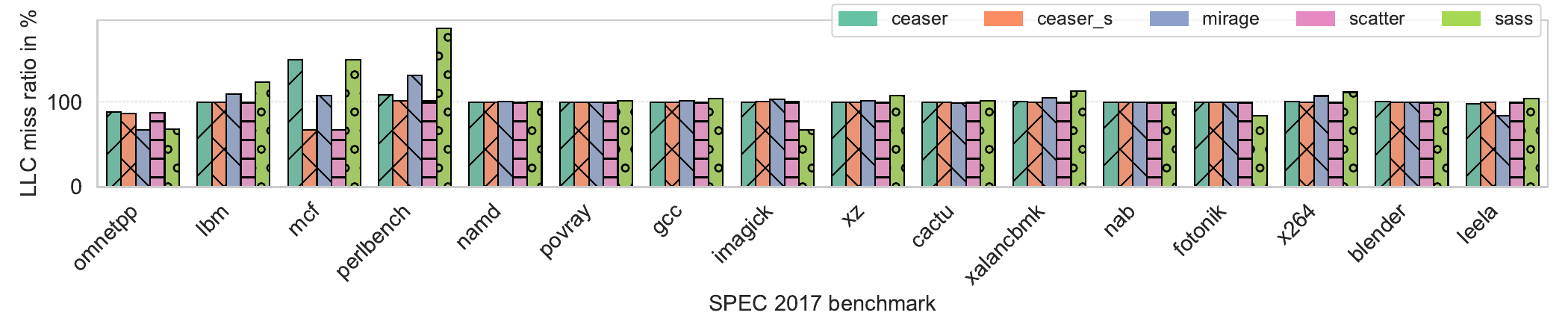}
    \caption{Performance evaluation of considered cache designs with \fiforp~replacement policy (normalized against baseline set-associative, and expressed as a $\%$). Performance statistics are averaged over $300$ copies of \spec~runs.}
    \label{fig:fiforp}
\end{figure*}

\end{document}